\documentclass[preprint2]{aastex6}

\usepackage{apjfonts}
\usepackage{graphicx}
\usepackage{natbib}
\usepackage{color}
\usepackage[title]{appendix}
\newcommand{\angstrom}{{\rm \AA}}
\newcommand{\lya}{Ly{$\alpha$}}
\newcommand{\NV}{N\,{\small V}\,$\lambda$1240}
\newcommand{\NVab}{N\,{\small V}\,$\lambda\lambda$1239,1243}
\newcommand{\NIV}{N\,{\small IV}]\,$\lambda$1486}
\newcommand{\NIVb}{N\,{\small IV}]\,$\lambda$1718}
\newcommand{\NIII}{N\,{\small III}]\,$\lambda$1750}
\newcommand{\CIV}{C\,{\small IV}\,$\lambda$1549}
\newcommand{\CIVab}{C\,{\small IV}\,$\lambda\lambda$1548,1551}
\newcommand{\HeII}{He\,{\small II}\,$\lambda$1640}

\newcommand{\SiII}{Si\,{\small II}\,$\lambda$1816}
\newcommand{\SiIII}{Si\,{\small III}]\,$\lambda$1892}
\newcommand{\AlIII}{Al\,{\small III}\,$\lambda$1857}
\newcommand{\CIII}{C\,{\small III}]\,$\lambda$1909}
\newcommand{\MgII}{Mg\,{\small II}\,$\lambda$2800}

\newcommand{\OIII}{O\,{\small III}]\,$\lambda$1663}
\newcommand{\hbeta}{H{$\beta$}}

\newcommand{\objfull}{SDSS~J120414.37$+$351800.5}
\newcommand{\obj}{SDSS~J1204+3518}

\slugcomment{Received 2018 February 15; revised 2018 March 15; accepted 2018 March 29; published 2018 May 16}

\begin{document}

\title{A Candidate Tidal Disruption Event in a Quasar at $z=2.359$ from Abundance Ratio Variability}

\shorttitle{Candidate TDE in $z>2$ Quasar}

\shortauthors{Liu et al.}

\author{Xin Liu\altaffilmark{1,2}, Alexander Dittmann\altaffilmark{1,3}, Yue Shen\altaffilmark{1,2,5}, Linhua Jiang\altaffilmark{4}}

\altaffiltext{1}{Department of Astronomy, University of Illinois at Urbana-Champaign, Urbana, IL 61801, USA; xinliuxl@illinois.edu}

\altaffiltext{2}{National Center for Supercomputing Applications, University of Illinois at Urbana-Champaign, 605 East Springfield Avenue, Champaign, IL 61820, USA}

\altaffiltext{3}{Department of Physics, University of Illinois at Urbana-Champaign, Urbana, IL 61820, USA}

\altaffiltext{4}{Kavli Institute for Astronomy and Astrophysics, Peking University, Beijing 100871, China}

\altaffiltext{5}{Alfred P. Sloan Foundation Fellow}

\begin{abstract}
A small fraction of quasars show an unusually high nitrogen-to-carbon ratio (N/C) in their spectra. These ``nitrogen-rich'' (N-rich) quasars are a long-standing puzzle because their interstellar medium implies stellar populations with abnormally high metallicities. It has recently been proposed that N-rich quasars may result from tidal disruption events (TDEs) of stars by supermassive black holes. The rapid enhancement of nitrogen and the depletion of carbon due to the carbon--nitrogen--oxygen cycle in supersolar mass stars could naturally produce high N/C. However, the TDE hypothesis predicts that the N/C should change with time, which has never hitherto been observed. Here we report the discovery of the first N-rich quasar with rapid N/C variability that could be caused by a TDE. Two spectra separated by 1.7 years (rest-frame) show that the \NIII /\CIII\ intensity ratio decayed by $\sim86$\%$\pm$14\% (1$\sigma$). Optical (rest-frame UV) light-curve and X-ray observations are qualitatively consistent with the TDE hypothesis; though, the time baseline falls short of a definitive proof. Putting the single-object discovery into context, statistical analyses of the $\sim80$ known N-rich quasars with high-quality archival spectra show evidence (at a 5$\sigma$ significance level) of a decrease in N/C on timescales of $>1$ year (rest-frame) and a constant level of ionization (indicated by the \CIII /\CIV\ intensity ratio). If confirmed, our results demonstrate the method of identifying TDE candidates in quasars via abundance ratio variability, opening a new window of TDE observations at high redshift ($z>2$) with upcoming large-scale time-domain spectroscopic surveys. 
\end{abstract}


\section{Introduction}\label{sec:intro}

Studies of the optical (rest-frame UV) spectra of high-redshift ($z>2$) quasars show that $\lesssim$1\% exhibit much stronger nitrogen emission (seen in \NV , \NIV , and/or \NIII ) compared to the collisionally excited lines of other heavy elements such as carbon \citep{Bentz2004,Bentz2004a,Jiang2008,Batra2014}. The high N/C is caused by significantly elevated nitrogen-to-carbon abundance ratios (e.g., by a factor of 10 in the prototypical case of Q0353$-$383) since physical conditions in the emission-line regions of N-rich quasars appear similar to ordinary quasars \citep{Shields1976}. The origin of N-rich quasars is generally attributed to unusually high metallicities \citep[e.g., $Z>5 Z_{\odot}$;][]{Dietrich2003,Nagao2006}. The high metallicities could result from either extreme global enrichment in the cores of giant elliptical galaxies \citep[][but see \citealt{Friaca1998,Romano2002} for arguments against this]{Hamann1993} or local enrichment in the central part of the quasar \citep{Collin1999,Wang2011}. However, it is difficult to generate a stellar population that has elevated nitrogen abundances unless the absolute metallicity is extremely high. 

TDEs have recently been proposed as an alternative, more natural explanation of the high N/C \citep{Kochanek2016a}. The requirement for unusually high metallicities is obviated since the N-rich phenomenon would be transient. The TDE population that may cause significant abundance anomalies would be dominated by supersolar mass stars since more massive stars are too rare and less massive stars would not have enough time for nuclear processing within the age of the universe.  Although such events are expected to be rare (e.g., $\sim$10\% of all TDEs), they could quickly increase N/C by factors of 3--10 \citep{Kochanek2016a,Gallegos-Garcia2018}. The disruption of even a single star could be enough to temporarily pollute the broad-line region (BLR) gas, assuming the BLR gas mass is on the order of $\gtrsim10^{-3} M_{\odot}$\footnote{The BLR gas mass is luminosity dependent and is likely higher in luminous quasars \citep{Baldwin2003a}.} \citep{Peterson1997}. Although significant differences exist between the rest-frame UV spectra of low-redshift TDEs and high-redshift N-rich quasars, unusually strong nitrogen emission is seen in all of the three optical TDEs that have UV spectra in low-redshift galaxies \citep{Cenko2016,Yang2017,Brown2017}.

To test the TDE hypothesis, we study the spectroscopic variability for a sample of 82 N-rich quasars compiled from the literature \citep{Jiang2008,Batra2014} that have high-quality (S/N$>10$ pixel$^{-1}$ in the spectral region of interest) archival spectra from the SDSS \citep{York2000} and SDSS-III/BOSS \citep{Dawson2013} surveys. Our main findings include:

\begin{enumerate}

\item Discovery of an $\sim86$\%$\pm$14\% (1$\sigma$) weakening of the \NIII /\CIII\ emission-line intensity ratio between 2005 and 2011 in a N-rich quasar. This is the most dramatic spectral change that has been observed in any N-rich quasar.

\item Demonstration that the observations of the decrease of the \NIII /\CIII\ emission-line intensity ratio can be explained by a TDE of a star as it gets torn apart by the supermassive black hole (SMBH) of the quasar, even though there are (perhaps necessary) differences between the candidate TDE (in a $z>2$ quasar) and the few well-studied TDEs known in the literature (mostly in low-redshift, inactive galactic nuclei).

\item Archival data on the X-ray and optical light curves of the quasar show a decrease in its apparent brightness that are qualitatively consistent with the TDE hypothesis, though the baseline falls short of a proof.

\item Statistical analyses of a parent N-rich quasar population provide evidence of a decrease in the  \NIII /\CIII\ emission-line intensity ratio on timescales of $>1$ year (rest-frame), whereas the \CIII /\CIV\ intensity ratio remains unchanged. This suggests that \obj\ is not just a statistical fluke. 

\item If confirmed, the TDE scenario obviates the problem of how and why there would be extremely nitrogen-enriched material in the nuclear regions of quasar host galaxies, an enrichment that has been difficult to explain by chemical evolution models for gas in galaxies. The disruption of an evolved star would naturally release N-rich gas in the vicinity of the quasar, at least temporarily.

\end{enumerate}

The rest of the paper is organized as follows. \S \ref{sec:main} presents our main result on the discovery of the quasar, \objfull\ (hereafter \obj ) with a spectroscopic redshift of $z=2.359$ \citep{schneider10}, as our best candidate for having significant N/C variability over $>1$ year (rest-frame) timescales. While the available observations do not provide a definitive proof, we show that they are consistent with the TDE hypothesis. Details on the data and methods are provided in \S \ref{sec:data}, which a more general reader may want to skip. \S \ref{sec:stat} presents statistical analyses that put the single-object discovery (\S \ref{sec:main}) in the context of the general N-rich quasar population. Finally, \S \ref{sec:discuss} discusses implications of our results and suggests directions for future work. We provide further checks on systematic uncertainties in the Appendices.

We adopt the \NIII /\CIII\ emission-line intensity ratio as an indicator of the N/C abundance ratio \citep{Batra2014,Yang2017}. The \NIII\ and \CIII\ lines have similar ionization potentials and critical densities. Detailed photoionization simulations have demonstrated that N$^{2+}$ and C$^{2+}$ are formed in the same volume of space and that the \NIII /\CIII\ ratio is a good indicator of N/C \citep{Yang2017} . Focusing on the ratio of emission lines with similar ionization levels and critical densities is crucial to separating changes in the abundance ratio from effects due to changes in the ionization level and/or gas density, because different emission lines may be formed in overlapping but different volumes of space in the BLRs \citep{Peterson1988}. We use the \CIII /\CIV\ emission-line intensity ratio to calibrate possible changes in the ionization level \citep{Shields1976,Baldwin2003}.

\begin{figure}
\centerline{
\includegraphics[width=0.5\textwidth]{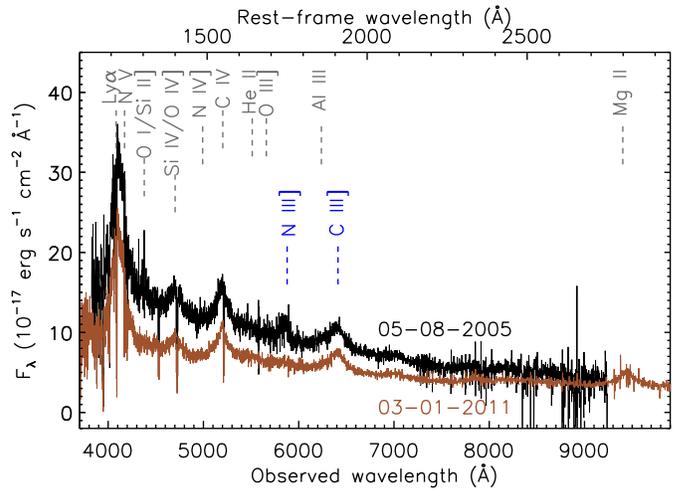}
} 
\caption{Optical (rest-frame ultraviolet) spectra of \obj\ (redshift $z=2.359$) reveal decreased \NIII /\CIII .
The earlier-epoch (later-epoch) spectrum from SDSS (BOSS) is shown in black (brown).
\label{fig:spec} }
\end{figure}

\section{Discovery of \obj\ as a Candidate TDE}\label{sec:main}

\subsection{\NIII /\CIII\ Intensity Ratio Decay}

\begin{figure}
\centerline{
\includegraphics[width=0.5\textwidth]{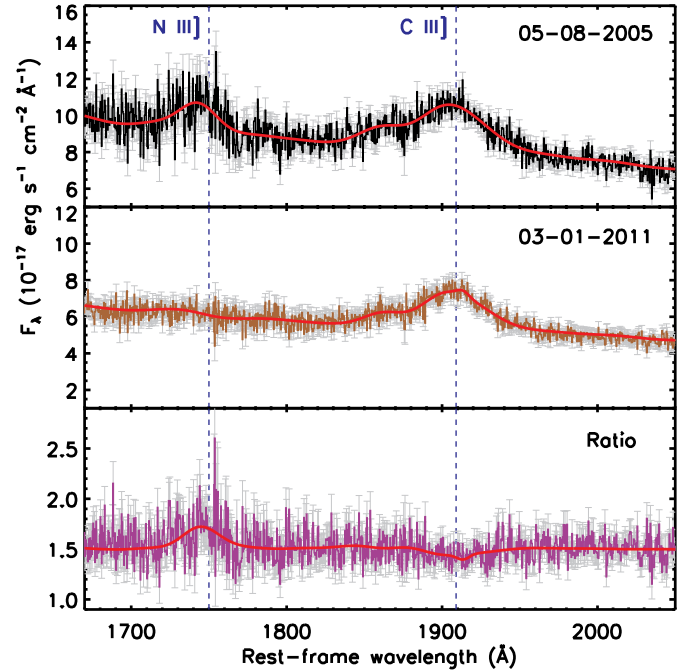}
} 
\caption{Spectral modeling of the \NIII\ and \CIII\ emission lines for \obj .
We show the flux density (thin black and brown curves), 1-$\sigma$ uncertainty (gray error bars), and our best-fit model (thick red curves) 
for the SDSS (top), BOSS (middle), and ratio (bottom) spectra.
\label{fig:spec_wfit}}
\end{figure}

Figure \ref{fig:spec} shows the two archival spectra of \obj . The SDSS spectrum was taken on 05-08-2005, whereas the BOSS spectrum was taken on 03-01-2011, i.e., 1.7 years (in the quasar's rest frame) after the SDSS spectrum. The flux intensity of the \NIII\ emission decayed significantly, whereas \CIII\ stayed constant within uncertainties (see Figure \ref{fig:spec_wfit} and Table \ref{tab:linedata} for details). We do not detect significant blueshift or redshift in \NIII . 

We do not detect significant \NIV\ or \HeII\ in either epoch (see \S \ref{subsec:otherlines} for discussion). The intensity of \NV\ seems to have decayed (Appendix A, Figure \ref{fig:linefit_nv}), though, it is highly uncertain due to the blending with Ly$\alpha$. We cannot robustly disentangle broad \NV\ absorption, if any, from emission due to blending and the intrinsically broad widths of the lines. 

The narrow absorption lines seen in the spectra include four intervening \CIVab\ doublet systems and two associated systems (a \CIVab\ doublet and a \NVab\ doublet) that have consistent redshifts (Appendix A, Figure \ref{fig:nal}). There are no clear and significant variations in the strengths or velocities of the narrow absorption lines. 

\subsection{Origin of the Rapid Abundance Ratio Variability}

The significant N/C decrease observed in \obj\ based on the \NIII /\CIII\ emission-line intensity ratio cannot be easily explained by extreme metallicities. The timescales for the stellar population evolution necessary for metallicity changes (in quasar host galaxy stellar core or in the outer self-gravitating part of the accretion disk) are much longer than a few years. 

Significant changes in the ionizing spectrum are unlikely given the similar ratios seen in the two-epoch spectra for other BLR emission lines such as the ionizing species of carbon \citep[e.g., the \CIII /\CIV\ intensity ratio is an indicator of the ionization level;][Table \ref{tab:linedata}]{Shields1976,Baldwin2003}. Furthermore, the \NIII /\CIII\ intensity ratio is sensitive to the N/C abundance ratio and is rather insensitive to the ionization parameter and the slope of the ionizing continuum \citep{Shields1976,Osmer1980,Baldwin2003,Yang2017}.  

Below we proceed with the hypothesis that the rapid N/C variability seen in \obj\ was caused by a candidate TDE of a star by the SMBH that powers the quasar.

\begin{table}
\centering
\begin{tabular}{lcc}
\hline\hline
~~~~~~Emission-line Measurements~~~~~~ & SDSS &  BOSS \\
\hline
\NIII\ Flux (10$^{-16}$ erg s$^{-1}$ cm$^{-2}$)          \dotfill(1) & 13.7$\pm$1.6 & 1.9$\pm$1.6  \\
\CIII\ Flux (10$^{-16}$ erg s$^{-1}$ cm$^{-2}$)         \dotfill(2) & 32.6$\pm$4.4 & 31.2$\pm$4.2   \\
\CIV\ Flux (10$^{-16}$ erg s$^{-1}$ cm$^{-2}$)         \dotfill(3) & 68.7$\pm$9.3 & 58.4$\pm$7.9   \\
\NIII /\CIII\                              \dotfill(4) & 0.42$\pm$0.08  & 0.06$\pm$0.06  \\ 
\CIII /\CIV\                              \dotfill(5) & 0.47$\pm$0.09 & 0.53$\pm$0.10  \\
\CIV\ FWHM (km s$^{-1}$)    \dotfill(6) & 6810$\pm$400 & 6640$\pm$470   \\
\MgII\ FWHM (km s$^{-1}$)    \dotfill(7) &       N/A               & 6040$\pm$290   \\
\NIII\ EW (\angstrom )          \dotfill(8) & 4.53$\pm$0.53 & 1.12$\pm$0.94  \\
\CIII\ EW (\angstrom )         \dotfill(9) & 12.5$\pm$1.7 & 18.5$\pm$2.5   \\
\CIV\ EW (\angstrom )         \dotfill(10) & 19.2$\pm$2.6 & 26.8$\pm$3.6   \\
\hline
\end{tabular}
\caption{Emission-line Measurements from the SDSS and BOSS Spectra.
Note. Lines (1)--(3): emission-line flux intensity.  
The \SiIII\ and \AlIII\ lines have been deblended and subtracted from the \CIII\ complex.
Lines (4), (5): emission-line intensity ratio. 
Lines (6), (7): FWHM of the emission line. The SDSS spectrum does not cover \MgII .
Lines (8)--(10): rest-frame equivalent width of the emission line.  
Errors quoted are 1-$\sigma$ uncertainties estimated from Monte Carlo simulations 
(for flux, FWHM, and EW measurements) or calculated from error propagation (for emission-line flux ratios).
}
\label{tab:linedata}
\end{table}

\subsection{Optical Light Curves}

\begin{figure}
\centerline{
\includegraphics[width=0.5\textwidth]{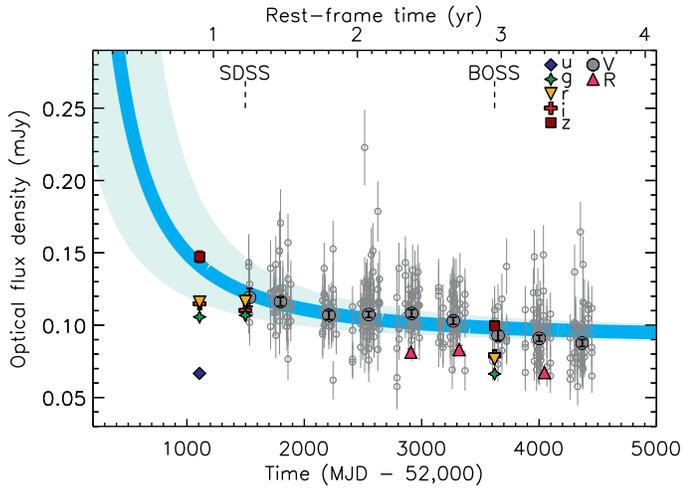}
} 
\caption{Optical light curves of the quasar \obj .
The data are from the CRTS (V band), 
PTF (R band), and SDSS ($u$, $g$, $r$, $i$, and $z$ band) surveys. 
We show both the individual (small open circles) and yearly inverse-variance-weighted mean (large filled circles) values for the CRTS data.
We only show the yearly inverse-variance-weighted mean (large filled upward-facing triangles) for the PTF data to avoid overcrowding.
The yearly inverse-variance-weighted mean values are centered on the median epoch of all the observations in a given year.
Error bars denote the 1$\sigma$ uncertainty in the mean flux density of a given year.
Also shown is the synthetic flux density in the corresponding SDSS filter calculated from the SDSS and BOSS spectra 
(the epoch of which is indicated by dashed lines). 
The cyan shading indicates the model of a $t^{-5/3}$ decay plus a constant background (with the light cyan shading representing 1$\sigma$ errors; see \S\S \ref{ap:olc_data}  \& \ref{ap:olc_fit} for details).
\label{fig:lc} }
\end{figure}

The optical (rest-frame UV) light curves of \obj\ are qualitatively consistent with the TDE hypothesis, though the baseline falls short of a proof. Figure \ref{fig:lc} shows the heterogenous set of archival data spanning photometric epochs from 2002 to 2013, encompassing the two spectroscopic epochs. There is evidence of a decaying component in the flux on top of a background of stochastic variability (at the level of $\sim$10\%) typically seen in optical quasars. As a toy model, we fit the light-curve data using the combination of a constant background plus a decaying component as $f\propto t^{\alpha}$ (see \S\S \ref{ap:olc_data} \& \ref{ap:olc_fit} for details) where $\alpha=-5/3$ is expected for TDEs for stars according to conventional TDE theory \citep{Rees1988}. A fit to the V-band data assuming $\alpha=-5/3$ and a constant background of 0.09 mJy yields a peak-luminosity date of 51,860$\pm$240 (MJD), which is $\sim$1.3 years (rest-frame) before the ``nitrogen-high'' state caught in the SDSS spectrum. The fit is highly uncertain since the peak-luminosity date is unknown. A model with $\alpha=-4$ \citep[expected for TDEs of partially disrupted stars;][]{Guillochon2013} fits the data equally well. 

Our toy models are for the purpose of illustration only and are not meant to be proof of a TDE.  Recent observations have shown that the conventional $t^{-5/3}$ profile provides a poor fit to the light curves of most low-redshift TDEs \citep{Arcavi2014,Holoien2014,Holoien2016,Holoien2016a,Brown2016,Brown2017a,Gezari2017}. A range of power-law profiles seem to fit various decline rates at different times after disruption.

While the available optical light-curve data are broadly consistent with our baseline model, the null hypothesis (i.e., purely stochastic quasar variability) cannot be easily ruled out given the limited temporal coverage and the poor photometric accuracy for most of the light-curve data. Nevertheless, the large variability amplitude observed ($\gtrsim30$\%) is unusual for purely stochastic quasar variability given its estimated high Eddington rate (\S \ref{ap:bhmass}), since high Eddington-rate quasars are unlikely to show large variability \citep{Rumbaugh2017}.  \S \ref{subsec:bright} discusses the brightness of the flare showing that it can be accommodated under the TDE hypothesis.

\subsection{X-Ray Observations}

Archival X-ray observations of \obj\ are broadly consistent with the TDE hypothesis. The X-ray flux decayed by $\sim$40\% over rest-frame $\sim$0.3 years (within 1 rest-frame year after the assumed peak luminosity date; \S \ref{ap:xray}), in agreement with our toy model at least qualitatively. Furthermore, An archival XMM-Newton observation taken on MJD=52820 suggests an extremely soft X-ray spectrum \citep[``photometric'' X-ray photon index $\Gamma_{\rm X}=4.1$ estimated from the slope between the luminosities at 1 and 5 keV;][]{Lusso2016}).  This is similar to X-ray flares in low-redshift TDEs \citep[$\Gamma_{\rm X}\gtrsim4$; e.g.,][]{Bade1996,Komossa1999a,Lin2015,Auchettl2017a,Auchettl2017}, but is significantly different from typical optical quasars with $\Gamma_{\rm X}\approx1.9$ \citep{Young2009}. We have analyzed the distribution of $\Gamma_{\rm X}$ using archival XMM-Newton observations for a control sample of ordinary quasars \citep{Lusso2016} that have similar redshift and luminosity to \obj . \obj\ is a $\sim$4$\sigma$ outlier in the distribution of $\Gamma_{\rm X}$. An archival Chandra observation taken on MJD=52476 also suggests a soft X-ray spectrum, though, the counts were too few for a robust measurement. \S \ref{ap:xray} presents details on the X-ray data and analysis.

\section{Details on the Data and Methods}\label{sec:data}

\subsection{Optical Spectroscopic Data and Analysis}\label{ap:data} 

\begin{figure*}
\centerline{
\includegraphics[width=0.7\textwidth]{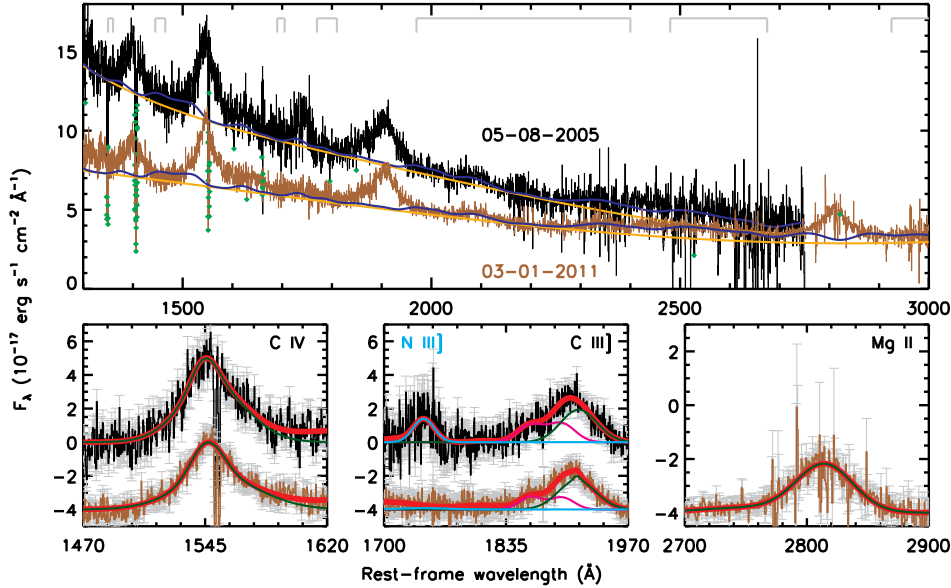}
} 
\caption{Spectral model fits for \obj .
The SDSS spectrum is shown in black and the BOSS spectrum is shown in brown.
Top panel shows the global fit to the pseudo-continuum (with the combined pseudo-continuum model in blue and the power-law component in orange).
The line segments in gray denote the wavelength windows adopted for the pseudo-continuum modeling.
The diamonds in green mark the pixels that have been rejected in the fit (from either absorption lines or bad pixels).
Bottom panels exhibit the model fits to the emission-line-only spectra (i.e., after subtracting the pseudo-continuum models) over the \CIV , \NIII --\CIII , and \MgII\ regions, respectively (with the combined emission-line model shown in red, the \CIV , \CIII , and \MgII\ component in dark green, the \AlIII\ and \SiIII\ complex in magenta, and the \NIII\ component in cyan). The error bars show the 1-$\sigma$ measurement error in the flux density. 
The BOSS emission-line-only spectrum is offset vertically by $-4\times10^{-17}$ erg s$^{-1}$ cm$^{-2}$ \angstrom $^{-1}$ for clarity.
\label{fig:linefit}
}
\end{figure*}

\obj\ is contained in the SDSS DR7 quasar catalog \citep{schneider10,shen11}. It has two spectra available from the SDSS DR13 data archive. The first epoch spectrum\footnote{http://dr13.sdss.org/sas/dr13/sdss/spectro/redux/26/spectra} (Plate = 2089, Fiber ID = 328, MJD = 53498) was from the SDSS-I/II survey \citep{York2000} and the second epoch\footnote{http://dr13.sdss.org/sas/dr13/eboss/spectro/redux/v5\_9\_0/spectra} (Plate = 4610, Fiber ID = 652, MJD = 55621) was taken by the BOSS spectrograph within the SDSS-III survey \citep{Dawson2013}. The SDSS spectrum covers the wavelength range of 3800--9200 \angstrom\ with a spectral resolution of $R=$1850--2200, whereas the BOSS spectrum covers 3650--10400 \angstrom\ with a similar spectral resolution. 

To measure the intensity ratio \NIII /\CIII\ (``NCR'' for short, defined in Equation \ref{eq:ncr}) as an indicator of the N/C abundance ratio, we fit spectral models to the observed spectra following the procedures as described in detail in \citet{Shen2012}. The model is a linear combination of a power-law continuum, a pseudo-continuum constructed from Fe\,II emission templates, and single or multiple Gaussians for the emission lines. As the uncertainties in the continuum model may induce subtle effects on measurements of the weak emission lines, we first perform a global fit to the emission-line free region to better quantify the continuum components.  After subtracting the continuum we then fit multiple Gaussian models to the emission lines around the \NIII\ and \CIII\ region locally. The \CIV\ region is also fit locally but separately to measure the intensity ratio \CIII /\CIV\ (``CCR'' for short, defined in Equation \ref{eq:ccr}) as an indicator of the ionization parameter or the hardness of the ionizing continuum. We also fit the \MgII\ region for virial black hole mass estimates from the BOSS spectrum.

More specifically, we model the \NIII\ line with a single Gaussian whose center and width are free parameters. We fit the wavelength range rest-frame 1700--1970 \angstrom . We model the \CIII\ line using up to two Gaussians. We adopt two Gaussians for the \SiIII\ and \AlIII\ lines which are often blended with the blue side of \CIII . To break degeneracy in decomposing the \CIII\ complex, we tie the centers of the two Gaussians for \CIII\ (i.e., the \CIII\ emission-line profile is constrained to be symmetric); we also tie the velocity offsets of \SiIII\ and \AlIII\ (relative to \CIII ) to the laboratory values. We also adopt three additional Gaussians for the other possibly detectable emission lines in the region \citep[\NIVb , Fe {\small II} UV 191 at 1786.7 \angstrom , and \SiII ;][]{Baldwin2003}. For the \CIV\ line, we fit the rest-frame wavelength range 1500--1700 \angstrom . We model the \CIV\ line using two Gaussians. We adopt four additional Gaussians for the narrow and broad components of \HeII\ and \OIII .  For the \MgII\ line (covered only in the BOSS spectrum but not in the SDSS spectrum), we fit the wavelength range rest-frame 2700--2790 \angstrom . We model the \MgII\ line using a combination of up to two Gaussians for the broad component and one Gaussian for the narrow component. We impose an upper limit of 1200 km s$^{-1}$ for the FWHM of the narrow lines. As an example, Figure \ref{fig:linefit} shows our spectral decomposition modeling for \obj . Figure \ref{fig:linefit_nv} shows the spectral fit around the \NV\ region, which is highly uncertain because of blending. Figure \ref{fig:nal} shows the narrow absorption line systems seen in the spectra of \obj . Appendix B discusses further checks on systematic uncertainties of the spectra.

\subsection{Optical Light-curve Data and Analysis}\label{ap:olc_data} 

\obj\ has available light-curve data from the Catalina Real-time Transient Survey \citep[CRTS;][$V$ band]{Drake2009}, the Palomar Transient Factory \citep[PTF;][$R$ band]{Law2009}, and the SDSS ($u$, $g$, $r$, $i$, and $z$ bands).  The SDSS measurements have the smallest photometric uncertainties, but there is only one photometric epoch and two subsequent spectroscopic epochs. We estimate the synthetic flux density in the corresponding SDSS filter calculated from convolving the SDSS and BOSS spectra with the filter throughput curves. We adopt the CRTS $V$-band magnitudes (converted to flux measurements) for fitting the light-curve model because they cover the longest time baseline that encompasses the two spectroscopic epochs. The PTF data do not provide more temporal coverage but serve as a double check for systematics. To mitigate the large photometric uncertainties associated with the CRTS and PTF data, we focus on the yearly inverse-variance-weighted-mean values in the analysis. We provide all the available photometry data in the literature in Table \ref{tab:photometry}. We have also checked the available LINEAR \citep{Sesar2011} data (re-calibrated to SDSS $r$ band) that extends the temporal coverage to earlier epochs, but \obj\ is close to the detection limit of the survey and its photometric uncertainties are too uncertain to be included in the analysis.

\begin{table*}
\centering
\begin{tabular}{lccccccc}
\hline\hline
        & $V$ (CRTS) & $R$ (PTF) & $u$ (SDSS) & $g$ (SDSS) & $r$ (SDSS) & $i$ (SDSS) & $z$ (SDSS) \\
MJD & (Vega mag) & (Vega mag) & (AB mag) & (AB mag) & (AB mag) & (AB mag) & (AB mag) \\
\hline
53,108      & ...                 & ... &  19.34 (0.03) & 18.84 (0.01) & 18.74 (0.01) & 18.75 (0.01) & 18.48 (0.03) \\
53,712.45 & 18.52 (0.14) & ... & ... & ... & ...   & ... & ... \\
53,712.46 & 18.89 (0.16) & ... & ... & ... & ...   & ... & ... \\	
53,712.47 & 18.66 (0.15) & ... & ... & ... & ...   & ... & ... \\	
53,712.48 & 18.67 (0.15) & ... & ... & ... & ...   & ... & ... \\	
53,527.20 & 18.77 (0.15) & ... & ... & ... & ...   & ... & ... \\	
53,527.21 & 18.71 (0.15) & ... & ... & ... & ...   & ... & ... \\	
53,527.22 & 18.49 (0.14) & ... & ... & ... & ...   & ... & ... \\	
53,527.23 & 18.51 (0.14) & ... & ... & ... & ...   & ... & ... \\	
53,534.16 & 18.86 (0.16) & ... & ... & ... & ...   & ... & ... \\	
53,534.17 & 18.70 (0.15) & ... & ... & ... & ...   & ... & ... \\	
\hline
\end{tabular}
\caption{Optical Photometry from the Literature.
This table is available in its entirety in machine-readable form.
\label{tab:photometry}}
\end{table*}

\subsection{Fits to the Optical Light Curve}\label{ap:olc_fit} 

The $V$-band corresponds to the rest-frame UV that best traces the mass accretion rate in TDEs. We fit the light curve with $C + f(t)$ where $C$ is a constant and $f(t)$ is a power law of the form 
\begin{equation}
f(t)=f_0\times (t+t_0)^{\alpha}.
\end{equation}
A fit to the nine data points assuming $\alpha=-5/3$ and a fixed constant background of $C=$0.09 mJy (estimated using the latest CRTS epochs) yields a best-fit model of $t_0 = -51,860\pm$240 (MJD) and $f_0 =7.5\pm1.5$ Jy ($\chi^2/\nu = 4.3$ for $\nu=$ seven degrees of freedom). Assuming $\alpha=-5/3$ and a positive constant background yields a model of $C=0.00$ mJy, $t_0 = -39,500\pm$1500 (MJD), and $f_0 =(1.0\pm0.2)\times10^3$ Jy ($\chi^2/\nu=1.4$ for $\nu=$ six degrees of freedom). While the latter model is statistically improved, we prefer the former model since the background quasar emission should be nonzero. The implied shorter evolutionary timescale is also more consistent with the variability timescale suggested by the two-epoch spectra. Allowing $\alpha$ to vary instead (assuming a fixed constant background of $C=$0.09 mJy) does not provide a statistically significant improvement to the fit (yielding $\chi^2/\nu = 4.0$ for $\nu=$ six degrees of freedom for a model of $\alpha=-3.97\pm0.05$).

\subsection{Estimation of the Candidate TDE Luminosity and Energy}\label{ap:energy}  

Assuming a constant background quasar emission of 0.09 mJy, we estimate that the $V$-band brightness of the candidate TDE is $\sim$0.04 mJy at MJD$\sim$53,000.  At the redshift of $z=2.359$, the implied absolute $V$-band magnitude is $M_V=-23.4$ mag, assuming a Friedmann-Robertson-Walker cosmology with $\Omega_m = 0.3$, $\Omega_{\Lambda} = 0.7$, and $h = 0.7$.  A K-correction of 3.1 mag has been applied assuming that the unknown SED follows a black body with $T=2\times10^4$ K which implies a rest-frame $V$-band apparent magnitude of $\sim$2.4 $\mu$Jy at MJD$\sim$53,000.  Assuming that the $V$-band bolometric correction is in the range of 1.8--60 \citep[appropriate for black body temperatures of $10^4$--5$\times10^4$ K as observed in known TDEs;][]{Tadhunter2017}, the bolometric luminosity of the candidate TDE is estimated to be in the range of $9.5\times10^{44}$---$3.2\times10^{46}$ erg s$^{-1}$ at MJD$\sim$53,000. The total energy released during the observed part of the candidate TDE flare (using a simple trapezium rule integration of the $V$-band light curve) was estimated as $\sim1\times10^{52}$--$4\times10^{53}$ erg. The implied total mass accreted by the black hole (from the candidate TDE only) is estimated as $\sim0.02$--$0.5M_{\odot}$, assuming an accretion disk efficiency of $\eta=0.42$ for a Kerr black hole \citep{Leloudas2016}. This is a lower limit because the peak luminosity is not covered by the available light-curve data.

\subsection{Estimation of the Black Hole Mass and Eddington Ratio}\label{ap:bhmass} 

We estimate the black hole mass using the single-epoch estimator assuming virialized motion in the broad-line region clouds \citep{Shen2013}. Spectral fit to the SDSS spectrum suggests a \CIV -based virial mass of $M_{\bullet}= 10^{9.6\pm0.1} M_{\odot}$ (statistical error only) using the calibrations of \citet{Vestergaard2006}. The BOSS spectrum suggests a \CIV -based virial mass of $M_{\bullet}= 10^{9.5\pm0.1} M_{\odot}$, or a \MgII -based virial mass estimate of $M_{\bullet}= 10^{9.3\pm0.1} M_{\odot}$, using the calibrations of \citet{Vestergaard2009} for \MgII . We adopt the BOSS estimates as our baseline values because they are more representative of the nitrogen-low (i.e., more steady accretion) state of the black hole.  \MgII-based masses are generally considered more reliable than \CIV-based masses \citep[e.g.,][]{Shen2013}, given the larger scatter between \CIV\ and \hbeta\ masses for high-redshift quasars \citep[e.g.,][]{Shen2012}. \CIV\ is more subject to nonvirial motion such as outflows. We estimate the Eddington ratio as $L_{{\rm bol}}/L_{{\rm Edd}}$, where the Eddington luminosity is $L_{{\rm Edd}}=M_{\bullet}/M_{\odot}\times10^{38}$ erg s$^{-1}$ and the bolometric luminosity $L_{{\rm bol}}$ is calculated from $L_{{\rm 1350}}$ using the bolometric correction BC$_{{\rm 1350}}$=3.81 from the composite quasar SED of \citet{Richards2006}.

\subsection{X-Ray Data and Analysis}\label{ap:xray} 

There are three X-ray observations of the field of \obj\ available in the public data archives. The ROSAT All Sky Survey \citep[RASS;][]{voges99,voges00} scan with the Position Sensitive Proportional Counter \citep{Pfeffermann1987} from 1990 November yielded a 3$\sigma$ upper limit of 0.02 counts s$^{-1}$, corresponding to an unabsorbed X-ray flux limit of $F_{{\rm 0.1-2.4\,keV}} < 2.9\times10^{-13}$ erg cm$^{-2}$ s$^{-1}$ \citep[assuming a $\Gamma_{{\rm X}} = 2$ power-law spectrum corrected for the Galactic column density\footnote{https://heasarc.gsfc.nasa.gov/cgi-bin/Tools/w3nh/w3nh.pl} of $1.7\times10^{20}$ cm$^{-2}$;][]{dickey90}. 

A Chandra observation of HS 1202$+$3538 (observation ID=3070) on 2002 July 21 (MJD=52476) serendipitously caught \obj\ at an off-axis angle of 10.7 arcmin. The observed-frame soft (0.5--2 keV) and hard (2--8 keV) counts in the 6.7 ks (4.5 ks corrected for vignetting) exposure were 16 and 9, respectively \citep{Gibson2008}. We derive an unabsorbed X-ray flux (corrected for the Galactic column density) of $F_{{\rm 0.5-10\,keV}}=2.6\times10^{-14}$ erg cm$^{-2}$ s$^{-1}$ for $\Gamma_{{\rm X}} = 4$, or $F_{{\rm 0.5-10\,keV}}=4.4\times10^{-14}$ erg cm$^{-2}$ s$^{-1}$ for $\Gamma_{{\rm X}} = 2$. The X-ray counts are too few for a robust measurement of $\Gamma_{\rm X}$. Nevertheless, the hardness ratio HR $\equiv ({\rm H}-{\rm S})/({\rm H}+{\rm S}) = -0.28$ suggests a soft X-ray spectrum, similar to the values observed in the X-ray flares of TDEs not long after peak luminosity \citep{Auchettl2017a}. 

Finally, \obj\ is contained in the 3XMM serendipitous source catalog DR5 \citep{Rosen2016} with DETID=101487424010032. The {\it XMM-Newton} European Photon Imaging Camera detected \obj\ on 2003 May 23 and 2003 June 30. The longer exposure taken on 2003 June 30 (observation ID=0148742401) covered \obj\ with a ``live'' exposure time of 20.1 ks at an off-axis angle of 10.0 arcmin. The X-ray photon index estimated from the slope between the luminosities at 1 and 5 keV is 4.09 with an X-ray S/N = 4.7 \citep{Lusso2016}.  This ``photometric'' X-ray photon index is not as reliable as the spectroscopic tracer, but provides a reasonable estimate of the X-ray spectral hardness. The derived unabsorbed X-ray flux (averaged over the two {\it XMM-Newton} epochs) is $F_{{\rm 0.5-10\,keV}}=1.5\times10^{-14}$ erg cm$^{-2}$ s$^{-1}$ for $\Gamma_{{\rm X}} = 4$, or $F_{{\rm 0.5-10\,keV}}=1.9\times10^{-14}$ erg cm$^{-2}$ s$^{-1}$ for $\Gamma_{{\rm X}} = 2$. For the Chandra and {\it XMM-Newton} detections, we adopt the X-ray estimates from assuming $\Gamma_{{\rm X}} = 4$ as our baseline values (shown in Figure \ref{fig:lc}), although the qualitative decaying trend still holds for $\Gamma_{{\rm X}} = 2$.

\subsection{Radio Loudness Upper Limit from FIRST}\label{ap:radio} 

\obj\ was covered by the FIRST survey footprint but was undetected with a 3$\sigma$ upper limit of $f^{{\rm obs}}_{{\rm 6\,cm}}<$ 0.381 mJy. Assuming that the radio flux follows a power law (i.e., $f_{\nu} \propto \nu^{\alpha}$), this translates into $f^{{\rm rest}}_{{\rm 6\,cm}}<$0.70 mJy for a spectral index $\alpha=-0.5$ \citep{Jiang2007}, or $<$1.0 mJy assuming $\alpha=-0.8$ \citep{Gibson2008}. Combined with the $f_{{\rm 2500}}$ measurement from the optical spectrum, the implied limit on the radio loudness parameter \citep{Kellermann1989}, i.e., $R\equiv f_{{\rm 6\,cm}}/f_{{\rm 2500}}$, is $<$4.9 ($<$7.8) using the SDSS (BOSS) spectrum assuming $\alpha=-0.5$, or $<$7.2 ($<$11) assuming $\alpha=-0.8$, which is marginally inconsistent with the radio-loud criterion R$>$10.

\section{Statistical Context for the Single-object Discovery}\label{sec:stat}

\begin{figure}
\centerline{
\includegraphics[width=0.47\textwidth]{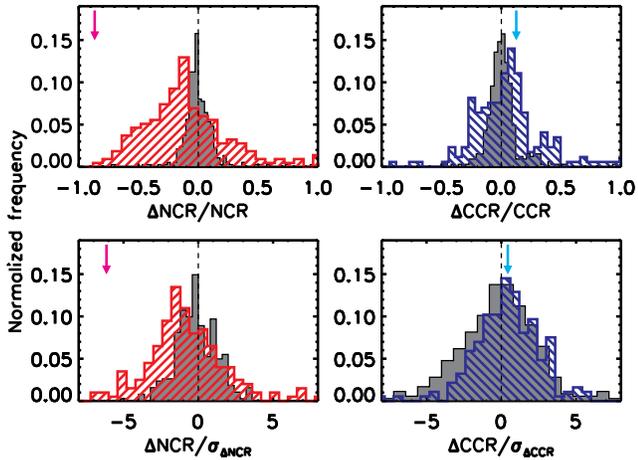}
} 
\caption{Probability distribution of intensity ratio variability reveals statistical evidence for \NIII /\CIII\ (``NCR'' for short) decreasing over $>$1 year timescales in N-rich quasars, whereas \CIII /\CIV\ (``CCR'' for short) remains unchanged.
Negative values denote decreasing with time.
The dashed histograms show the long-term (rest-frame $\Delta t>1$ year) variability. 
The gray histograms represent the short-term (rest-frame $\Delta t<1$ month) variability that is consistent with being purely due to statistical noise.
The upper panels show the fractional (or relative) variability of the emission-line intensity ratio whereas the lower panels display the S/N of the variability (i.e., variability normalized by measurement uncertainty accounting for both statistical and systematic errors). Arrows indicate the measurements for \obj .
See \S \ref{subsec:stat_result} for details.
\label{fig:distribution}
}
\end{figure}

\subsection{Sample Selection}\label{subsec:stat_sample}

We start with a sample of 311 N-rich quasars by combining two N-rich quasar catalogs in the literature \citep{Jiang2008,Batra2014}. The first catalog \citep{Jiang2008} contains 293 quasars at $1.7<z<4.0$ with strong \NIV\ or \NIII\ emission lines (rest-frame EW$>3$ \angstrom ).  The second catalog contains 43 quasars (of which 25 overlap with the first catalog) at $2.29<z<3.61$ that have the strongest \NIV\ and \NIII\ lines in addition to strong \NV\ lines. Both catalogs were selected from the fourth edition of the SDSS Quasar Catalog \citep{Schneider2007} based on the SDSS fifth data release \citep{SDSSDR5}.

To study spectroscopic variability for these N-rich quasars, we analyze all the available high-quality spectrum pairs (with median S/N $>$10 pixel$^{-1}$ over rest-frame 1700--2000 \angstrom ) in the SDSS thirteenth data release \citep{SDSSCollaboration2016} for a sample of 82 unique quasars. We visually examine the ratio spectrum (similar to that shown in the lower panel of Figure \ref{fig:spec_wfit} but over the entire relevant spectral range) for each spectrum pair; \obj\ was identified as our best candidate for having significant N/C variability (i.e., the ratio spectrum changed significantly over the \NIII\ region but stayed constant over the \CIII\ region). There are 193 high-quality spectrum pairs with rest-frame time separations $>$1 year (for 78 of the 82 unique N-rich quasars; the other 4 only have high S/N repeat spectra separated by $<$ 1 year). This ``long-term'' group serves as our parent sample for estimating the fraction of N-rich quasars that show significant N/C variability over $\gtrsim$ 1year timescales. To calibrate measurement uncertainties, we also define a control sample of a ``short-term'' group containing 355 high-quality spectrum pairs for N-rich quasars with rest-frame time separations $<$1 month. The median time baseline $<\Delta t>$ in the ``long-term'' (``short-term'') group is 2.7 (0.034) years (rest frame).

\subsection{Statistical Analysis}\label{subsec:stat_result} 

\begin{table}
\centering
\begin{tabular}{lcccccc}
\hline\hline
~~~~~~~~~~~~~~~~Sample & \multicolumn{2}{c}{Short-term ($<$1 mon)} & & \multicolumn{2}{c}{Long-term ($>$1 years)} &  \\
Measurements & \multicolumn{2}{c}{($<\Delta t>$ = 0.4 mon)} & & \multicolumn{2}{c}{($<\Delta t>$ = 2.7 years)} &  \\
    \cline{2-3} \cline{5-6} \\
Statistics & Median & SD & & Median & SD & P$_{{\rm null}}$ \\
\hline
$\Delta$NCR/NCR\dotfill      & $-$0.01$\pm$0.01 & 0.14 & & $-$0.12$\pm$0.08 & 1.2   & 10$^{-21}$  \\
$\Delta$CCR/CCR\dotfill      & 0.00$\pm$0.01      & 0.18 & & 0.06$\pm$0.06      & 0.79 & 10$^{-8}$   \\
$\Delta$NCR/$\sigma_{\Delta{\rm NCR}}$\dotfill   & $-$0.2$\pm$0.1 & 1.4 & & $-$1.0$\pm$0.2 & 2.8  & 10$^{-10}$   \\
$\Delta$CCR/$\sigma_{\Delta{\rm CCR}}$\dotfill   &  0.0$\pm$0.2     & 3.1  &  & 0.5$\pm$0.2     & 2.2  &  0.05   \\
\hline
\end{tabular}
\caption{Emission-line Intensity Ratio Variability Statistics.
Note. NCR stands for the \NIII /\CIII\ intensity ratio and CCR stands for the \CIII /\CIV\ intensity ratio.
The difference is defined as the later epoch minus the earlier epoch so that negative values mean a decrease in the intensity ratios over time.
The long-term (short-term) sample contains 193 (355) pairs of high-quality spectra for N-rich quasars.
SD: standard deviation. 
P$_{{\rm null}}$: null probability from the Kolmogorov--Smirnov test 
of the long-term ($>$1 years) sample compared against the short-term ($<$1 month) sample. 
See \S \ref{subsec:stat_result} for details.
\label{tab:stat}}
\end{table}

Figure \ref{fig:distribution} (upper left panel) shows the distribution of the fractional variability in the \NIII /\CIII\ emission-line intensity ratio (``NCR'' for short). We define the fractional variability as 
\begin{equation}\label{eq:ncr}
\Delta {\rm NCR/NCR} \equiv ({\rm N/C}_2 - {\rm N/C}_1)/({\rm N/C}_1),
\end{equation}
where N/C$_1$ is from the earlier epoch and N/C$_2$ is from the later epoch. Negative values mean a decrease in N/C over time.  Monte Carlo simulations suggest that the observed distribution of the short-term NCR variability is consistent with being purely induced by statistical noise. The dispersion in $\Delta$NCR/NCR for the long-term pairs is significantly larger than that in the short-term pairs (with the standard deviation (SD) being 1.2 for the long-term sample compared to 0.14 for the short-term sample; Table \ref{tab:stat}). This much larger scatter is most likely caused by systematic uncertainties associated with continuum modeling, which is larger in the long-term pairs due to increased quasar variability \citep[e.g.,][]{MacLeod2010,Morganson2014}. 

Figure \ref{fig:distribution} also shows the effective ``S/N'' of the NCR and CCR variation (lower panels), i.e., emission-line intensity ratio variation normalized by measurement uncertainties (accounting for both statistical and systematic errors). The dispersion in $\Delta$NCR/$\sigma_{\Delta{\rm NCR}}$ for the long-term pairs is more similar to that in the short-term pairs after the normalization (with SD being 2.8 for the long-term sample compared to 1.4 for the short-term sample), that accounts for differences in the systematic uncertainties.  The long-term sample shows a 5$\sigma$ detection that the median value is nonzero ($-1.0\pm0.2$, where 0.2 is the 1-$\sigma$ error in the median, estimated as SD$/\sqrt{N}$, where $N$ is the number of data points; Table \ref{tab:stat}). Furthermore, its probability distribution is significantly different from that expected from pure measurement noise as characterized by the short-term pairs. Kolmogorov--Smirnov test shows that the probability that the two distributions are drawn from the same sample $P_{{\rm null}}$ is $10^{-10}$ (Table \ref{tab:stat}). Compared to the control sample of short-term pairs, the long-term pairs show a net negative $\Delta$NCR/$\sigma_{\Delta{\rm NCR}}$, providing statistical evidence that NCR decreases with time in N-rich quasars. \obj\ stands out among those that show the most significant decreases in the NCR.  We have checked the other positive and negative value tails seen in the $\Delta$NCR/NCR distribution, which are largely caused by systematic uncertainties from our emission-line measurements and do not show significant variability in the flux-ratio spectrum. 

Figure \ref{fig:distribution} (upper right panel) also shows that the \CIII /\CIV\ emission-line intensity ratio (``CCR'' for short) variation. Similarly to the NCR variation, it is defined as
\begin{equation}\label{eq:ccr}
\Delta {\rm CCR/CCR} \equiv ({\rm C3/C4}_2 - {\rm C3/C4}_1)/({\rm C3/C4}_1),
\end{equation}
where C3/C4$_1$ is from the earlier epoch and C3/C4$_2$ is from the later epoch.  The long-term CCR variation, on the other hand, is more similar to the short-term CCR variation. While the dispersion is still larger for the long-term CCR variation in terms of $\Delta$CCR/CCR (SD being 0.79 for the long-term compared to 0.18 for the short-term sample; Table \ref{tab:stat}), their difference is smaller than that in the dispersion of $\Delta$NCR/NCR (a factor of $\sim4$ difference rather than $\sim9$). This is understandable considering that stronger lines (i.e., \CIV ) are less affected by systematic uncertainties in the continuum modeling. In terms of $\Delta$CCR/$\sigma_{\Delta{\rm CCR}}$, i.e., after normalizing the measurement uncertainties, the median value of the long-term sample is consistent with being zero within 2.5 $\sigma$ ($0.5\pm0.2$; Table \ref{tab:stat}); the 2.5$\sigma$ positive deviation from zero is most likely explained by a noise-induced bias, as we discuss below. Furthermore, the long-term and short-term distributions are statistically identical (with $P_{{\rm null}}$ being 0.05; Table \ref{tab:stat}). \obj\ does not show significant CCR variation. Table \ref{tab:stat} summarizes the statistical properties of the NCR and CCR variability measurements.

\begin{figure}
\centerline{
\includegraphics[width=0.47\textwidth]{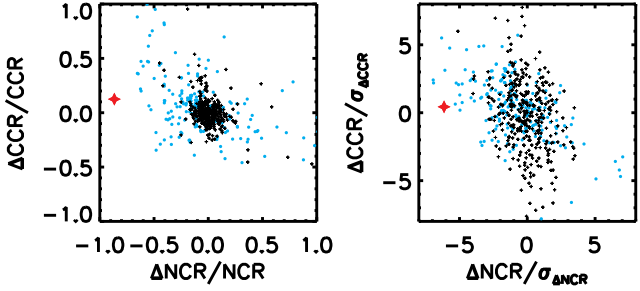}
} 
\caption{Test on NCR versus CCR.
While CCR appears to be increasing as NCR decreases in the long-term sample (shown with cyan dots), 
this apparent trend is also seen in the short-term sample (shown with black crosses), 
which is most likely caused by the inter-correlation between the NCR and CCR terms given measurement errors. 
The two parameters NCR and CCR are not independent from each other since the \CIII\ emission-line intensity is 
included in both terms. Considering measurement errors, a larger C III] (caused by noise) will result in 
a smaller NCR and a larger CCR.  \obj\ is shown with a red star in both panels.
\label{fig:ncr_ccr}
}
\end{figure}
\begin{figure}
\center{
\includegraphics[width=0.47\textwidth]{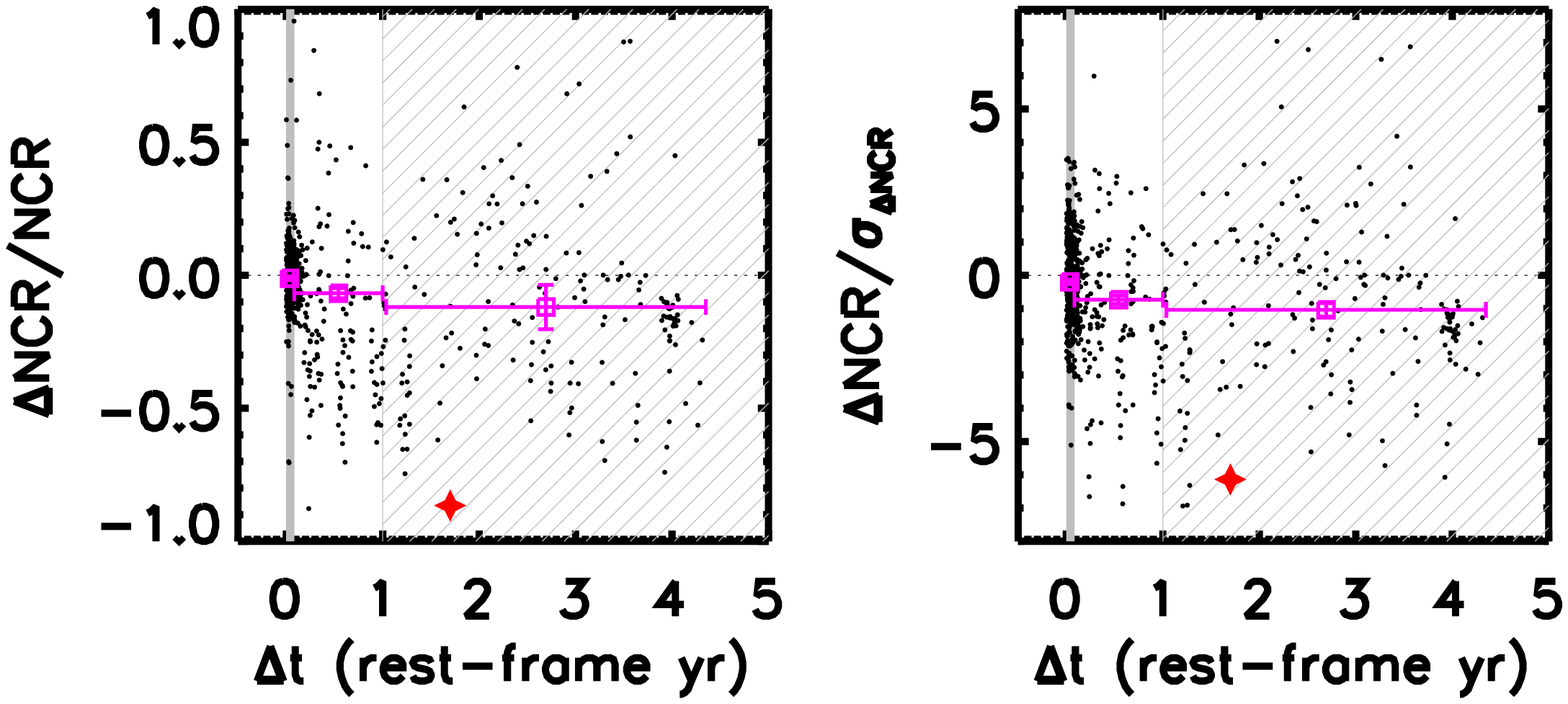} \\
\includegraphics[width=0.47\textwidth]{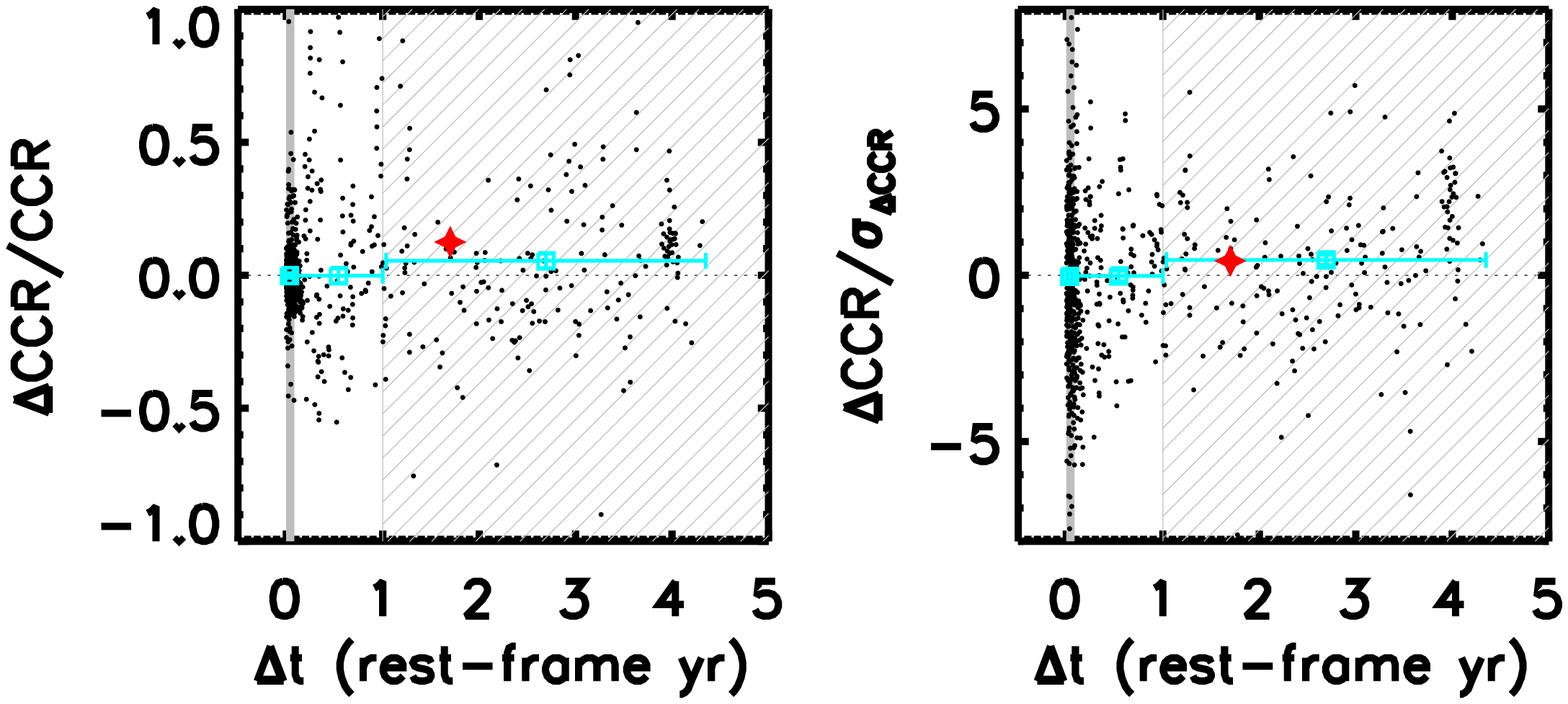}
} 
\caption{Examining the time dependence of intensity ratio variability.
Negative values denote decreasing with time.
Individual data points are shown with small black dots whereas large open squares denote median values in a given bin of time separation.
Error bars denote uncertainties in the medians.
The gray shaded region denotes the short-term (rest-frame $\Delta t<1$ month) sample, whereas the hatched region encloses the long-term (rest-frame $\Delta t>1$ years) sample. 
The left column shows the fractional (or relative) variability of the emission-line intensity ratio whereas the right column displays the S/N of the variability (i.e., variability normalized by measurement uncertainty). Stars indicate the measurements for \obj .
\label{fig:var_dt}
}
\end{figure}

Figure \ref{fig:ncr_ccr} examines whether there is any general trend between $\Delta$NCR/NCR and $\Delta$CCR/CCR. Although $\Delta$CCR/CCR appears to be increasing as $\Delta$NCR/NCR decreases in the long-term sample (shown with cyan dots), this apparent trend is also seen in the short-term sample (shown with black crosses), which is most likely caused by the correlation between the NCR and CCR terms due to measurement errors. The apparent trend arises because NCR and CCR are not independent from each other since the \CIII\ emission-line flux intensity is included in both terms. Considering measurement errors, a stronger \CIII\ (caused by noise) will lead to a smaller NCR and a larger CCR, resulting in an apparent anti-correlation between the NCR and CCR. This also likely explains the 2.5$\sigma$ positive deviation from zero in the median value of $\Delta$CCR/$\sigma_{\Delta{\rm CCR}}$ for the long-term sample, in which systematic uncertainty from continuum modeling adds to the error, resulting in more bias in the long-term sample than in the short-term sample because of increased quasar variability.

Figure \ref{fig:var_dt} examines the potential time dependence of the intensity ratio variability. We show both the individual data points (small dots) and the median values in a given bin of time separation (large open squares with error bars denoting uncertainties in the median values) to help assess the general trend. While there is some hint of the variability increasing as a function of time in support of the TDE hypothesis, we cannot draw a firm conclusion in view of the limited dynamic range in time separation and the large scatter in the intensity ratio variability measurements.

\section{Discussion and Future Work}\label{sec:discuss}

\subsection{Puzzles in the Emission Lines}\label{subsec:otherlines}

The absence of evidence for strong variability in the \NIV\ and \HeII\ lines is puzzling, but does not rule out the TDE hypothesis. In the only three known TDEs with available UV spectra \citep{Cenko2016,Yang2017,Brown2017}, \NIV\ (\HeII ) is not always detected or relatively weak in some epochs, whereas \NIII\ is always detected. So empirically it is possible to see \NIII\ but not \NIV\ (\HeII ) emission in TDEs, which may be related to the different ionizing potentials of the lines. In addition, a practical complication is that \NIV\ is on the extended blue wing of \CIV\ for \obj\ and may be buried underneath, especially if it is weak. For \HeII , another possibility is that if the disrupted star were about to ascend the giant branch, most helium would be sequestered in the dense helium core, which would not be disrupted by the more massive black hole (unlike in low-redshift TDEs around smaller black holes). Finally, our selection is based on the \NIII /\CIII\ ratio variability so that we may have missed systems that show stronger \NIV\ and/or \HeII\ variability. 

While bearing some resemblances to N-rich quasar spectra, none of the three TDEs with UV spectra exhibit \CIII\ or \MgII\ emission features. Their \CIV\ is also much weaker than expected. The origins of these discrepancies are still open to debate. They might be due to differences in the physical conditions in the gas and on the shape of the ionizing continuum.  For example, the absence of \MgII\ could result from photoionization by the extremely hot continuum, which may be transient in nature as the continuum temperature eventually cools down \citep{Cenko2016}. The absence of \CIII\ could be due to collisional deexcitation if the gas density is above critical, $n=10^{9.5}$ cm$^{-3}$ \citep{osterbrock89}. 

\subsection{\obj\ Was Not Too Bright for a TDE}\label{subsec:bright}

The brightness of the candidate TDE of $\sim$0.04 mJy at MJD$\sim$53,000 implies an absolute $V$-band magnitude of $-23.4$ mag at the redshift of the quasar. This is only $\sim0.6$ mag fainter than the peak luminosity of ASASSN-15lh \citep{Leloudas2016}, the brightest TDE candidate reported to date (see \citealt{Dong2016,Godoy-Rivera2017}, however, for an alternative explanation for the nature of the source as a supernova).  This high apparent brightness seems difficult to explain under the TDE hypothesis, but it is perhaps understandable for the following reasons. First, the observed $V$-band samples the rest-frame UV that is close to the peak of the spectral energy distribution (SED) of the candidate TDE, resulting in a large K-correction at the redshift of the quasar. This is dramatically different from local TDEs, in which the $V$ band only samples the Rayleigh-Jeans tail of the SED. 

Second, the high luminosity of the candidate TDE could be explained by the large BH mass of the quasar and its likely rapid spin (see \S\S \ref{ap:bhmass} \& \ref{subsec:discuss_mass_spin} for details). In addition, TDEs in AGN hosts may have higher radiative efficiencies than those in inactive galactic nuclei potentially due to interaction of the stellar debris with the preexisting accretion disk \citep{Blanchard2017}. However, the estimated efficiency does not strain the Eddington limit even at the projected peak-luminosity date, which is highly uncertain. The total energy released would require the accretion of a mass of $\gtrsim$0.02--0.5$M_{\odot}$ (see \S \ref{ap:energy} for details). 

Finally, our sample is most likely highly biased to the most luminous events because: (i) by selection we are sensitive only to the high-redshift (i.e., $z>1.7$ for the relevant UV lines to be covered in the observed optical spectra) population, and (ii) the SDSS is not a particularly deep survey.

\subsection{SMBH Mass and Spin of \obj }\label{subsec:discuss_mass_spin}

Analysis of the BOSS spectrum suggests that the quasar is powered by an SMBH with a virial mass estimate of 
\begin{equation}
{\rm log}~ (M_{\bullet}/M_{\odot})= 9.3\pm0.1\pm0.5~{\rm dex,}
\end{equation}
where 0.1 (0.5) dex is the 1$\sigma$ statistical (systematic) error \citep{Shen2013}. The implied Eddington ratio is $\sim$0.005--0.16 (for the candidate TDE flare component only, depending on the unknown SED; it is $\sim0.13$ for the background quasar emission; see \S \ref{ap:bhmass} for details) around MJD of 53,000 (i.e., $\sim1$ rest-frame years post the peak-luminosity date), consistent with an initial sub-Eddington phase predicted by conventional TDE theory for massive BHs \citep{Rees1988}.  Assuming a time lag of 
\begin{equation}
t_f\sim1.1 (M_{\bullet}/10^8 M_{\odot})^{1/2}r_{\star}^{3/2}m_{\star}^{-1}~{\rm years,}
\end{equation}
where $r_{\star}\equiv R_{\star}/R_{\odot}$ and $m_{\star}\equiv M_{\star}/M_{\odot}$, between disruption and infall of the tightest-bound debris for a solar-type star, a black hole mass of $M_{\bullet}= 10^{9.3} M_{\odot}$ would yield a Newtonian estimate of the disruption date of $\sim46,000$ (MJD), i.e., $\sim5$ rest-frame years before the peak-luminosity date. A relativistic estimate of the disruption date would be $\sim50,000$ (MJD), i.e., $\sim2$ rest-frame yr before the peak-luminosity date \citep{Leloudas2016}. 

The gravitational radius increases with black hole mass at a higher rate than the tidal radius $R_{{\rm T}}$ does (i.e., $R_{g} \propto M_{\bullet}$ whereas $R_{{\rm T}} \propto M_{\bullet}^{1/3}$). Therefore, stars can be tidally disrupted before being swallowed whole into the horizon only if the black hole is less massive than the Hills mass $M_{{\rm Hills}}$, where 
\begin{equation}
M_{{\rm Hills}}\approx 10^8 M_{\odot}r_{\star}^{3/2}m_{\star}^{-1/2}
\end{equation}
for non-spinning Schwarzschild black holes \citep{Hills1975}. $M_{{\rm Hills}}$ can be up to an order of magnitude larger for stars on optimal orbits around rapidly spinning Kerr black holes \citep{Leloudas2016}. The virial mass estimate of \obj\ suggests that a spinning Kerr black hole is required for all allowed masses. A main-sequence supersolar mass star on prograde equatorial orbits can be disrupted by a maximally spinning Kerr black hole with $M_{\bullet} = 10^{8.9} M_{\odot}$, or a moderately spinning Kerr black hole with $M_{\bullet}= 10^{8.5} M_{\odot}$. If the BH mass is in the higher end in the estimated range, then a giant star is required. 

The rapid N/C variability, on the other hand, implies that the evolutionary timescale is not much longer than a few years. Assuming that it is driven by the circularization luminosity \citep[but see, e.g.,][for an alternative explanation]{Jiangyf2016} that evolves on the fallback timescale $t_f$, which is comparable to the viscous timescale in the accretion disk for massive black holes, the rapid evolution suggests that the disrupted is a main-sequence star rather than a giant. Future near-infrared spectroscopy of the Balmer lines and/or reverberation mapping observations are needed to help better constrain the black hole mass for \obj .

\subsection{SMBH Mass and Spin of N-rich versus Normal Quasars}\label{subsec:nrich_mass_spin}

A significant difference between N-rich and normal optical quasars is that N-rich quasars have systematically narrower widths of \CIV\ \citep{Jiang2008,Batra2014}. This implies, under standard assumptions, that N-rich quasars have less massive black holes than the normal quasar population. This is at odds with the more massive host galaxies implied by the high metallicity hypothesis given the galaxy mass-metallicity relation \citep{tremonti04,erb06a,Maiolino2008} and the SMBH-host mass correlation \citep{ferrarese00,gebhardt00,tremaine02,KormendyHo2013,Graham2016}. In contrast, the TDE scenario would provide a natural explanation for the narrower widths of \CIV\ seen in N-rich quasars since less massive black holes are more likely to disrupt stars. 

N-rich quasars also have a significantly higher radio-loud fraction compared to normal optical quasars \citep{Jiang2008}. The origin of this difference is unknown. If the radio loudness is physically related to the black hole spin \citep{Blandford1977,sikora07}, the TDE hypothesis may also explain the high radio-loud fraction in N-rich quasars, since black holes with larger spins are more prone to disrupting stars. \obj\ was undetected by the FIRST 1.4 GHz survey \citep{white97} whose upper limit was marginally inconsistent with being radio loud (see \S \ref{ap:radio} for details).

\subsection{Disrupting Giants versus Main-sequence Stars}

\obj\ represents the first case of a quasar with significant N/C variability over yearly timescales. Our analysis on a large sample of N-rich quasars shows statistical evidence that N/C decays over time, while the ionization level remains unchanged (\S \ref{sec:stat}). However, similar to the case of Q0353$-$383 \citep{Osmer1980,Baldwin2003}, the typical decay rate estimated in the average SDSS N-rich quasars (median $\Delta$NCR/NCR$=-12\pm$8\% in $\sim$2.7 rest-frame years, or $\sim -4\pm$3\% yr$^{-1}$; Table \ref{tab:stat}) is much less than that seen in \obj\ ($\Delta$NCR/NCR$=-86\pm$14\% in 1.7 rest-frame years, or approximately $-50\pm$8\% yr$^{-1}$; Table \ref{tab:linedata}). This could be explained in the TDE scenario if most N-rich quasars were caused by TDEs of giants rather than main-sequence stars, since TDEs of giants would evolve more slowly \citep{MacLeod2012}. We cannot rule out the possibility that only some but not all of the N-rich quasars are caused by TDEs. Testing whether most N-rich quasars are due to TDEs of giants would require high S/N spectroscopic follow-up observations on much longer timescales (i.e., decades and even centuries) rather than being probed by current surveys.  

While the lifetimes of giant stars are much shorter than those of main-sequence stars (and therefore TDEs of giants should be much less frequent), it is still possible for the N-rich population to be dominated by TDEs of giants considering the disruption constraint given the large black hole masses of high-redshift quasars (i.e., main-sequence stars are more likely to be swallowed whole rather than being disrupted unless the black hole mass is small enough and/or the spin is high), in contrast to the demographics seen in local, inactive galaxies with smaller black holes.

\subsection{Implications on the TDE Rate in Quasar Hosts}\label{ap:tderate} 

We estimate the implied TDE rate in quasars assuming that (i) \obj\ is a TDE of a main-sequence star, and (ii) nearly all of the rest of N-rich quasars are TDEs of giants. The total order-of-magnitude rate is estimated as 
\begin{equation}
r=r_{{\rm MS}}+r_{{\rm G}} = f_{{\rm N-rich}}\times\bigg(\frac{f_{{\rm MS}}}{{\rm 2~yr}} + \frac{f_{{\rm G}}}{{\rm 10^3~yr}}\bigg)\sim10^{-4} ~{\rm yr}^{-1}~{\rm galaxy}^{-1},
\end{equation}
where $f_{{\rm N-rich}}\lesssim10^{-2}$, $f_{{\rm MS}}\sim\frac{1}{78}$, and $0<f_{{\rm G}}\lesssim\frac{77}{78}$ ($f_{{\rm G}}\sim\frac{77}{78}$ assuming all N-rich quasars are TDEs). The evolution timescale (which goes into the denominator of the rate estimate) is much longer for TDEs of giants \citep[][here assumed to be $\sim10^3$ years; assuming it is $\sim10^2$ years instead would make $r_{{\rm G}}$ comparable to $r_{{\rm MS}}$ but still not an order-of-magnitude larger]{MacLeod2012} so that their contribution to the estimated rate is small (or similar if $f_{{\rm G}}$ is close to unity) compared to that by \obj\ alone.

Under the TDE scenario, our systematic search implies a much higher TDE rate in quasars than in normal galaxies. The fraction of N-rich quasars is $\lesssim1\%$, and we found one definitive case \obj\ out of a parent sample of 78 N-rich quasars in total within a time window of $\sim 2$ years (rest-frame). Thus the implied TDE rate is $\sim10^{-4}$ yr$^{-1}$ galaxy$^{-1}$. This is $\sim$100 times higher than the expected TDE rate at $M_{\bullet}\sim10^9 M_{\odot}$, which is $\sim10^{-6}$ yr$^{-1}$ galaxy$^{-1}$ for supersolar mass stars \citep{Kochanek2016}. Similarly, the observed TDE rate in post-starburst galaxies is $\sim100$ times higher than normal galaxies \citep{Arcavi2014}. This is perhaps not a coincidence since both quasars and post-starburst activities may be associated with recent merger events that would greatly increase the number of stars in low angular momentum orbits, fueling TDEs. 

\subsection{Future Directions}

Our findings provide motivation for future research programs on N-rich quasars, which have been enigmatic and hard-to-understand objects until now. Despite their rarity, we show that N-rich quasars may be important links for understanding how SMBHs disrupt stars.

Examples of future research programs include:

\begin{enumerate}

\item Finding additional cases of TDEs of stars in quasars to establish their frequency of incidence and learn more about their astrophysics.

\item Learning more about the physics of the encounters of stars with SMBHs.

\item Provide improved knowledge of the evolutionary nature and distributions of stars near SMBHs. For example, time scales for disruption of giant stars are longer than for main-sequence stars. 

\end{enumerate}

The abundance ratio variability offers a potentially new method for identifying TDEs, complementary to the traditional method based on X-ray and/or UV/optical flux variability. In particular, it may open a new window of discovering TDEs at significantly higher redshifts ($z>2$) than previous work. In comparison, the majority of known TDEs are at low redshifts ($z<0.3$), and the current redshift record holder is Swift J2058+05 at $z=1.1853$\footnote{https://tde.space/} \citep{Cenko2012}. This has implications for the current efforts to use TDEs to study supermassive black holes. Its potential may be better realized with the upcoming large-scale time-domain spectroscopic surveys such as the SDSS-V \citep{Kollmeier2017} and MSE \citep{McConnachie2016} projects.

%------------------------------------------------------------------------------
\acknowledgments

We thank Z. {Ivezi{\'c}} and J.S. Stuart for assistance with the LINEAR data, A. Barth, B. Fields, Y. Jiang, Z. Li, and P. Maksym for discussions, and an anonymous referee for a prompt and careful report that improved the paper. X.L. acknowledges a Center for Advanced Study Beckman fellowship. Y.S. acknowledges support from the Alfred P. Sloan Foundation and NSF grant 1715579. L.J. acknowledges support from the National Key Program for Science and Technology Research and Development (2016YFA0400703), and from the National Science Foundation of China (11533001). This work makes extensive use of SDSS-I/II and SDSS-III/IV data (http://www.sdss.org/ and http://www.sdss3.org/). 

The SDSS and BOSS spectra are publicly available at http://dr13.sdss.org. The optical light curves from the SDSS, CRTS, PTF, and LINEAR are publicly available at http://dr13.sdss.org, http://crts.caltech.edu/, http://www.ptf.caltech.edu/, and http://skydot.lanl.gov/. The X-ray data from the Chandra and XMM-Newton telescopes are publicly available at http://cda.harvard.edu/ (ObsID=3070) and http://xmm2.esac.esa.int/ (Observation ID=0148742401). The best-fit spectral models to the SDSS and BOSS spectra are available upon reasonable request.

Facilities: Sloan

%\clearpage

\appendix

\begin{appendices}

\section{Spectral Modeling for \NV\ and Narrow Absorption Line Systems}\label{ap:morefigs} 

\begin{figure}
\centerline{
\includegraphics[width=0.5\textwidth]{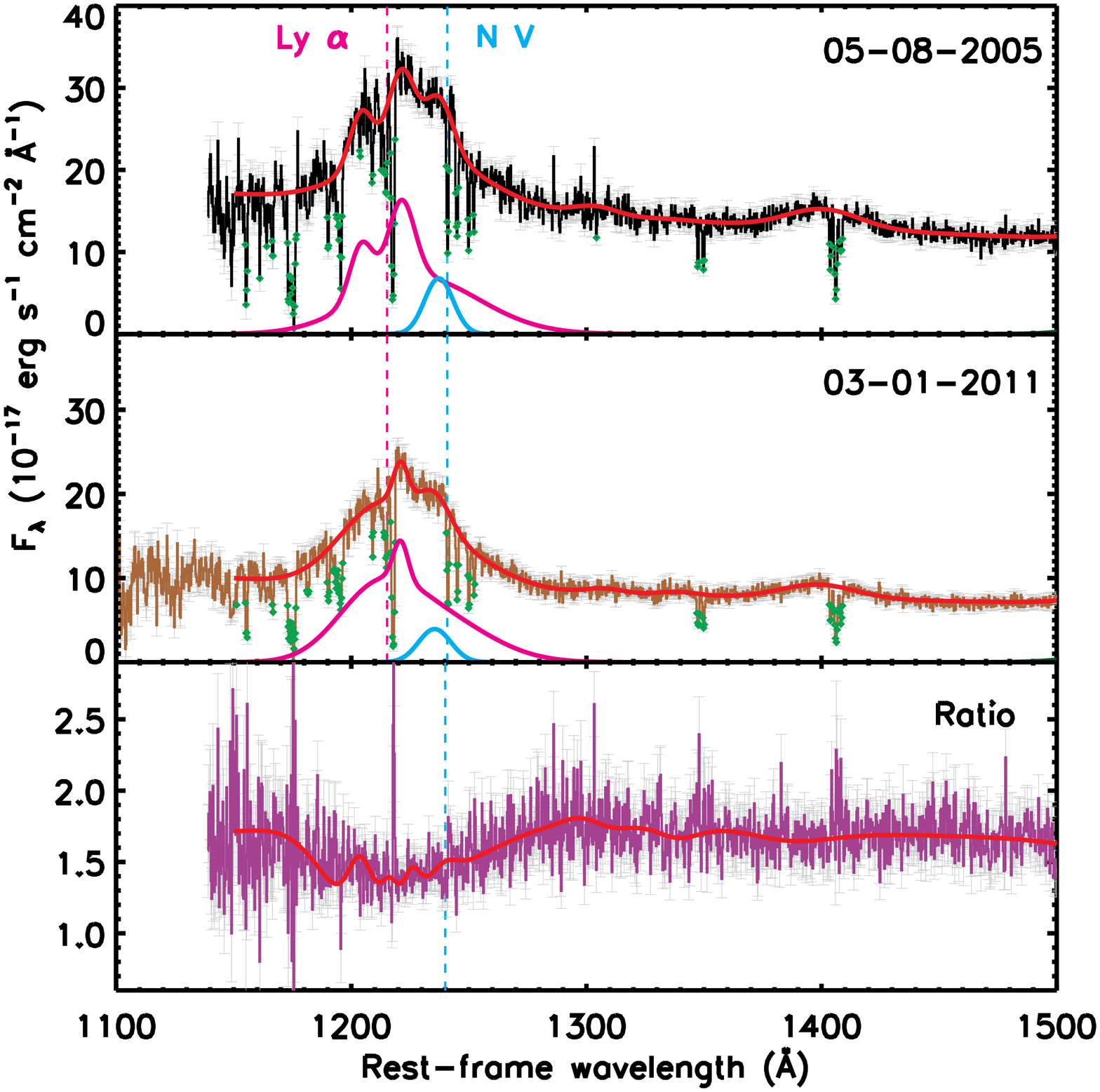}
} 
\caption{Spectral modeling for \obj\ around the \NV\ region.
Shown are the flux density (thin black and brown curves), 1$\sigma$ uncertainty (gray error bars), and our best-fit model (thick red curves) 
for the SDSS (top), BOSS (middle), and ratio (bottom) spectrum.
In the top and middle panels, we also plot models for the \lya\ (magenta) and \NV\ (cyan) complex. 
Although \NV\ seems to have decayed at face value, it is highly uncertain due to blending with \lya\ and possible contamination from broad absorption. 
Our model does not attempt to decouple emission from possible broad absorption because it is ambiguous given the broad widths of the lines.
The diamonds in green mark the pixels from narrow absorption lines that have been rejected in the fit.
\label{fig:linefit_nv}
}
\end{figure}

\begin{figure*}
\centerline{
\includegraphics[width=0.95\textwidth]{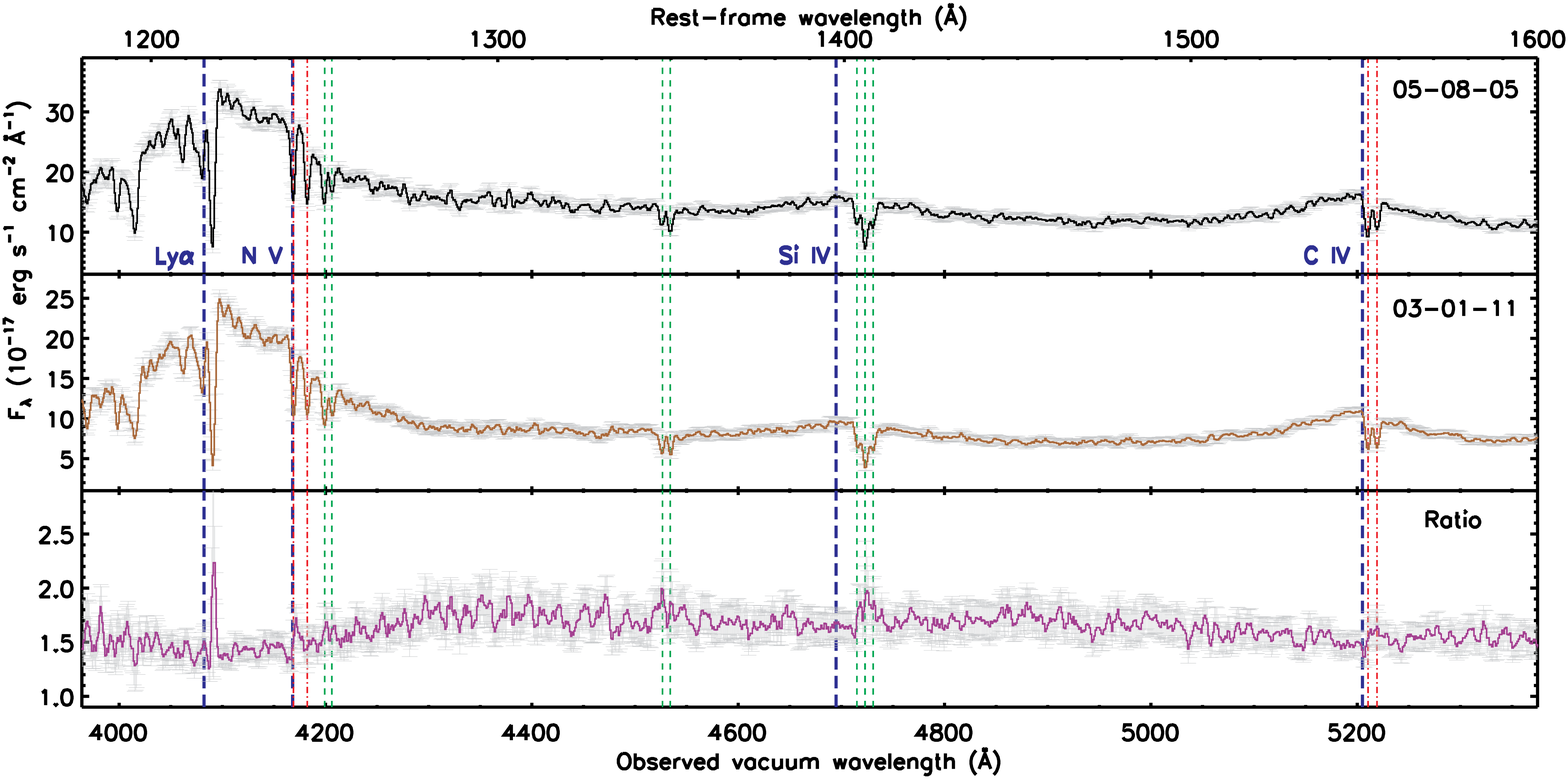}
} 
\caption{Narrow absorption line systems seen in the spectra of \obj .
Top (middle) panel shows the earlier-epoch (later-epoch) spectrum from SDSS (BOSS). Bottom panel shows the ratio spectrum. 
Blue thick long dashed lines mark the expected central wavelengths of the emission lines at the redshift of \obj .
Red thin dotted--dashed lines indicate narrow associated absorbers (a \CIVab\ doublet and a \NVab\ doublet) redshifted by $\sim550$ km s$^{-1}$ to the quasar systemic redshift.
Green thin dashed lines denote four intervening narrow \CIVab\ doublet absorbers (at the redshifts of 2.0506, 2.0455, 1.9238, and 1.7123) caused by the intergalactic medium along our line of sight to the quasar.
All spectra have been smoothed by a 4-pixel boxcar for display purposes. 
Error bars indicate 1$\sigma$ uncertainties pixel$^{-1}$ before smoothing.
\label{fig:nal}
}
\end{figure*}

Figure \ref{fig:linefit_nv} shows our spectral modeling for \obj\ around the \NV\ region. Figure \ref{fig:nal} shows narrow absorption line systems seen in the spectra of \obj .


\section{Systematic Uncertainties}\label{ap:systematics} 

We use a Monte Carlo approach \citep{Shen2012} to estimate measurement uncertainties, taking into account both statistical uncertainties due to flux errors and systematic uncertainties due to ambiguities in decomposing multiple model components.  As we discuss below, we have carried out extensive tests to validate the significance of the \NIII\ emission in the SDSS spectrum.  The detection is significant at the $>5 \sigma$ confidence level and is unlikely to be explained by noise (counting both systematic and statistical uncertainties).  

\begin{figure*}
\centerline{
\includegraphics[width=0.48\textwidth]{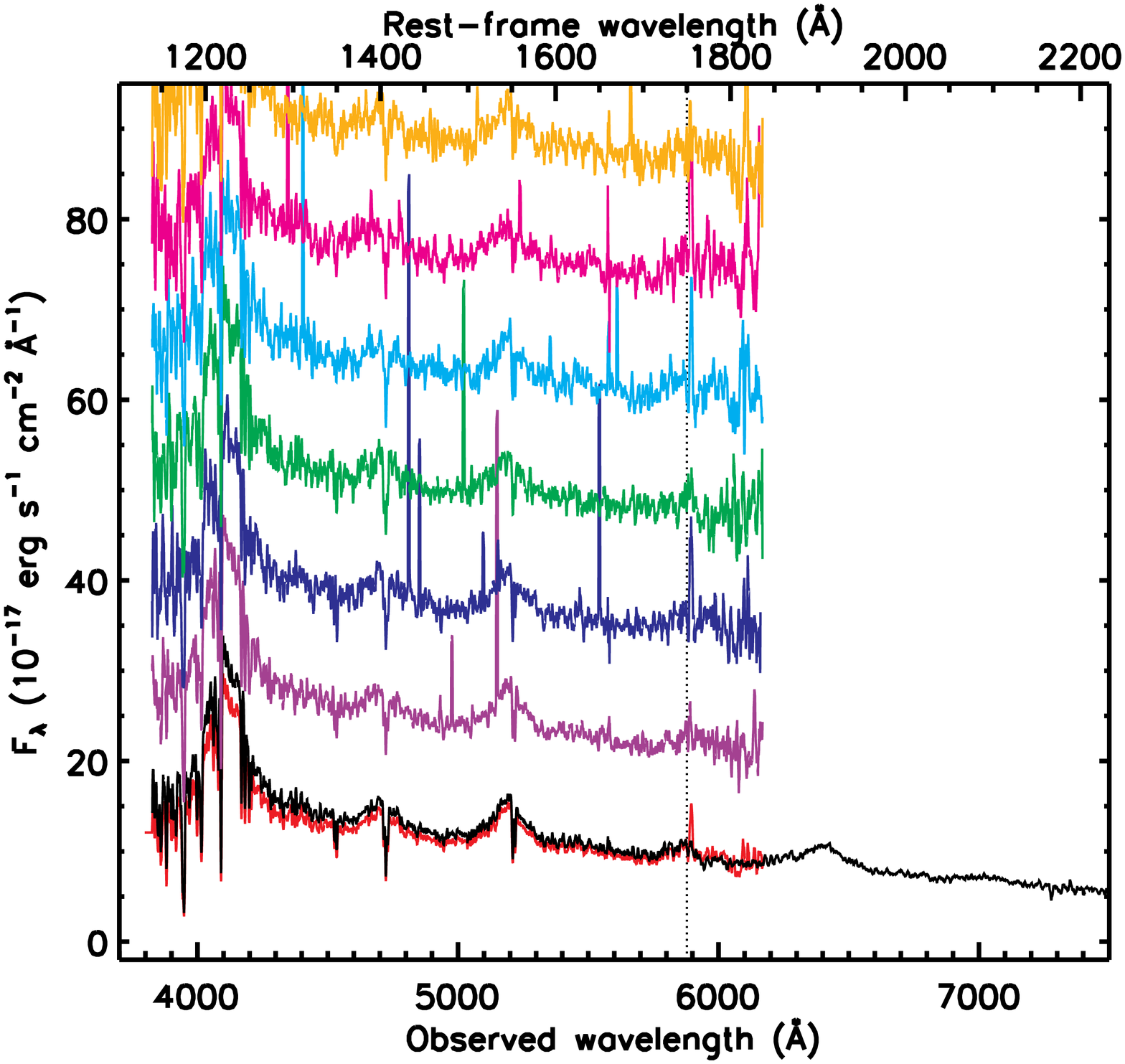}
\includegraphics[width=0.48\textwidth]{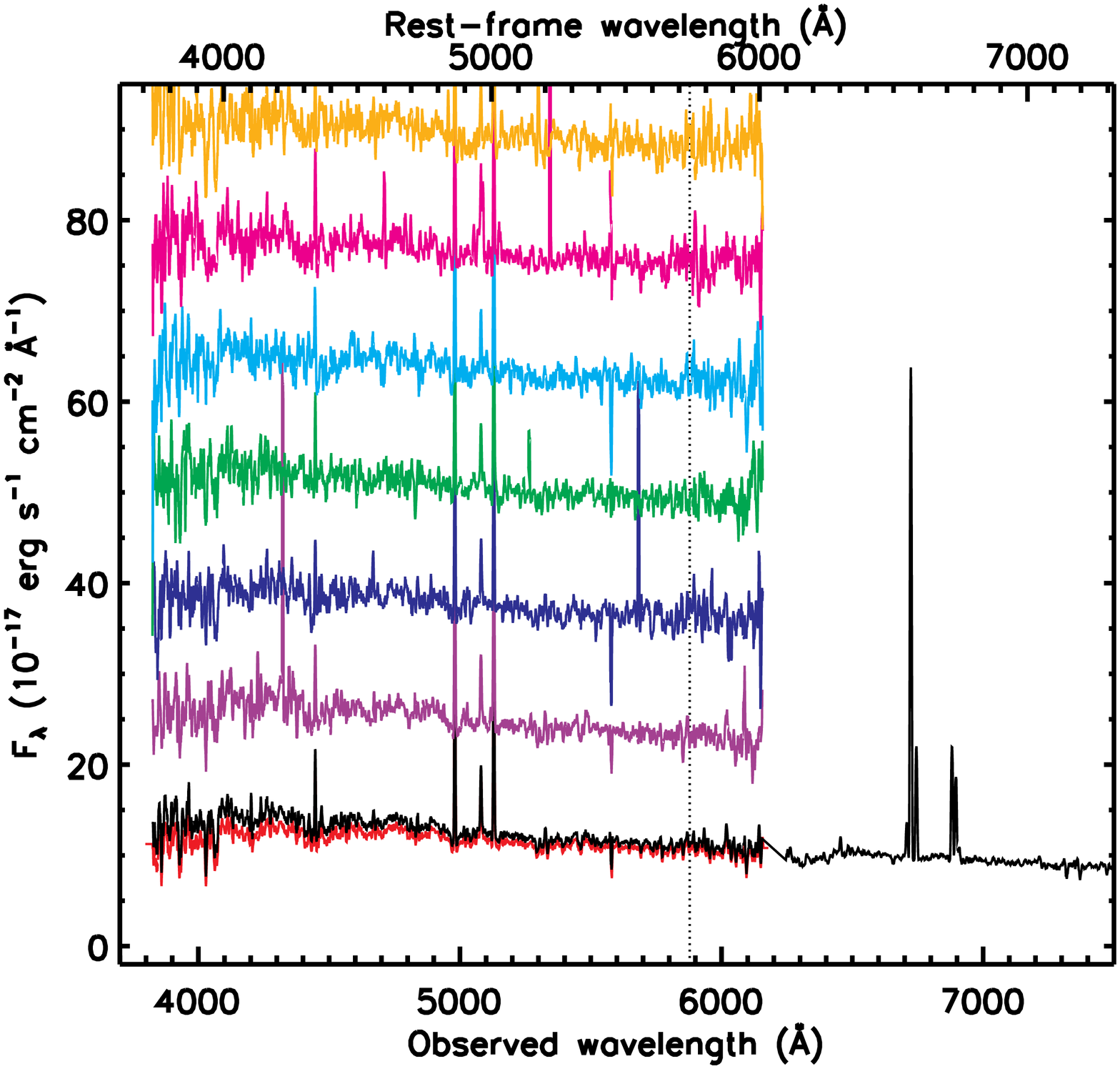}
} 
\caption{Individual exposure spectrum from the blue arm of the SDSS spectrograph for \obj\ (left panel) and 
a comparison object (a star-forming galaxy at $z=0.0242$; right panel) observed on the same plate but with a different fiber (ID=320).
The final co-added and blue--red stitched SDSS spectrum is shown in black, whereas the co-added blue spectrum is shown in red. 
The small difference between the co-added blue and the blue--red stitched spectra (seen for both \obj\ and the comparison object) 
is due to different scaling in the combination procedure for image distortion correction.
Individual blue spectra (shown in various colors) are offset vertically for display purposes and are ordered with a decreasing median S/N pixel$^{-1}$ over 5700--6100 \angstrom\ from bottom to top.
The expected wavelength of the \NIII\ line is marked by the dotted line. 
The high-frequency spike (seen in the co-added blue spectrum and some of the individual exposure spectra but not in the final blue--red stitched and co-add spectrum) is sky line residual from Na I $\lambda\lambda$5890,5896.
Despite the increased noise at the dichroic transition around redward of 6000 \angstrom\ (seen in the relatively featureless spectrum for the comparison object), \NIII\ is detected in each and every individual spectrum for \obj\ and is most clearly detected in the exposure with the best S/N (shown in purple).
All spectra have been smoothed by a 4-pixel boxcar for better clarity of the broad emission-line features.
\label{fig:test_noise}
}
\end{figure*}

First, the SDSS and BOSS spectra were both co-added, combining six consecutive individual 15-minute exposures. We have checked the individual exposures (Figure \ref{fig:test_noise}) to verify our emission-line measurement based on the co-added spectra.  Second, the \NIII\ feature is close to the dichroic edge of the SDSS spectrograph. To quantify possible systematic effects due to the dichroic edge, we have co-added and examined the individual SDSS spectra for the blue and red sides, separately. We have used the inverse-variance weighted mean to properly co-add the individual spectra. We have properly accounted for and rejected bad pixels.  

Figure \ref{fig:test_noise} (left panel) shows the co-added and individual blue SDSS spectrum. \NIII\ is detected in both the blue co-add and each and every of the individual exposures, although it is much more noisy in the individual spectra (so are the other emission lines).  In particular, it is most clearly detected in the exposure with the highest S/N (shown in purple; the individual exposure spectrum is ordered with increasing median S/N over the 5700--6100 \angstrom\ from top to bottom). We focus on the blue side here because it extends to the observed wavelengths of $\sim$6150 \angstrom , which covers the entire \NIII\ feature. The red side only partially covers the \NIII\ feature.  The fact that \NIII\ is detected in the blue-only spectrum argues against its origin as artifacts due to stitching together the red and blue spectra. Our final quoted detection significance is based on the co-added, blue--red stitched spectrum, because co-adding the red and blue spectra enhanced the S/N (Figure \ref{fig:test_noise}), which is particularly helpful for detecting weak lines. 

\begin{figure*}
\centerline{
\includegraphics[width=0.45\textwidth]{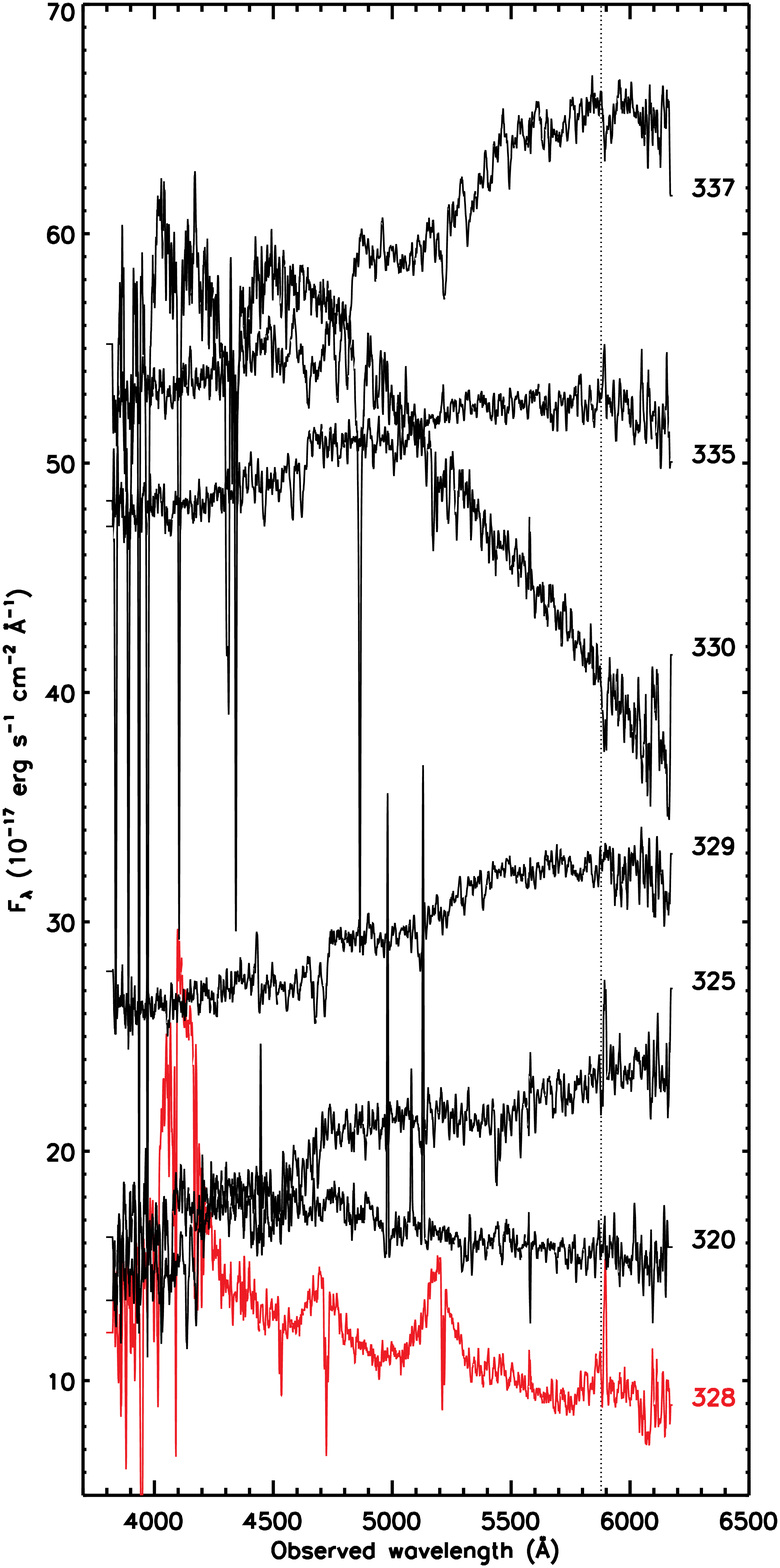}
\includegraphics[width=0.45\textwidth]{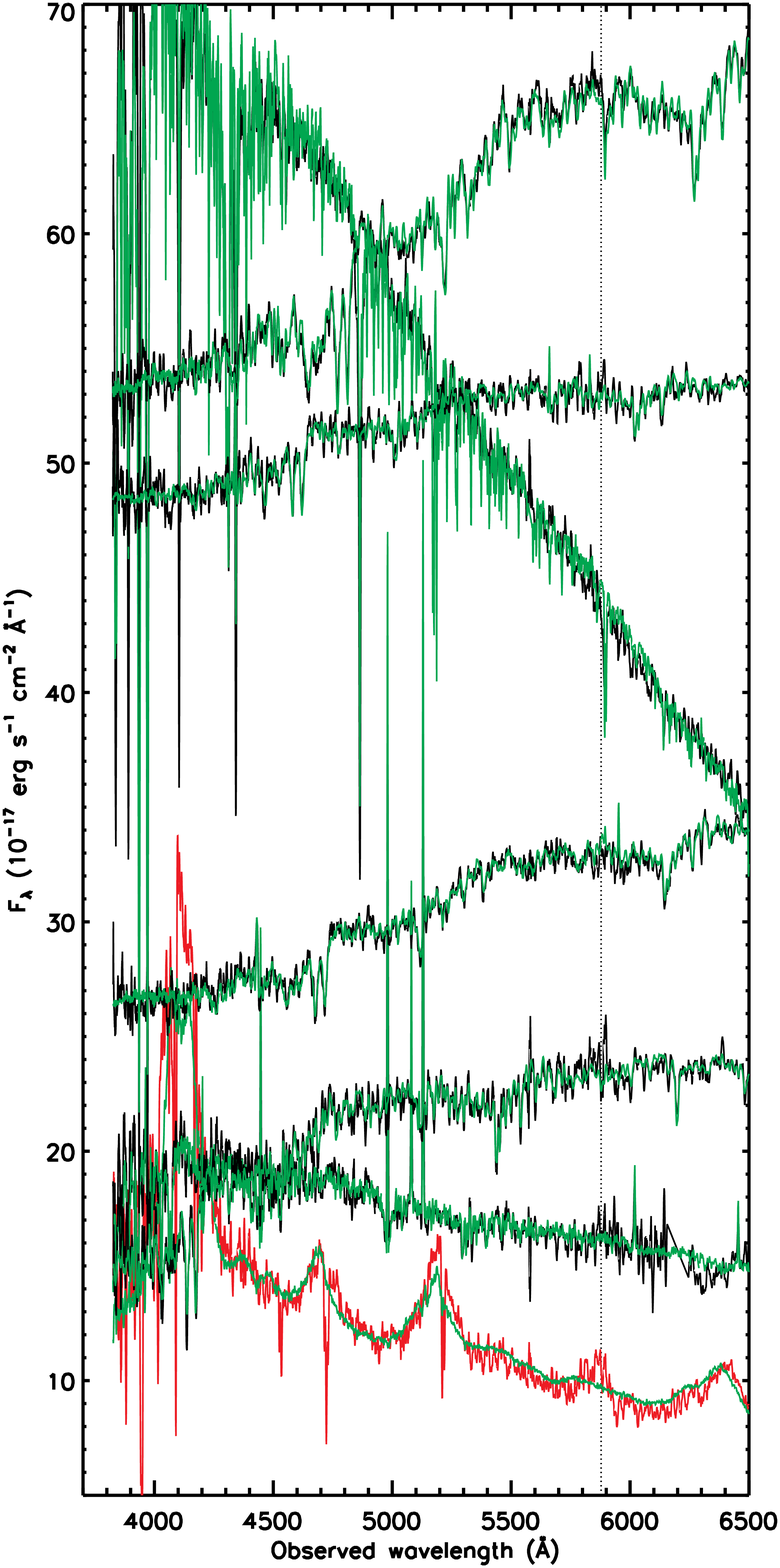}
} 
\caption{Systematics check for instrumental problems.
(Left) Co-added blue-side spectrum for \obj\ (shown in red) compared to 
spectra taken by other fibers (shown in black; fiber numbers are labeled next to each spectrum) on the same plate in the same night.
(Right) Co-added and blue--red stitched spectrum for \obj\ (shown in red) compared to those for other fiber spectra (shown in black).
Overplotted are the best-fit models (shown in green) from the SDSS automatic spectroscopic pipeline.
Only \obj\ shows significant residual emission around the expected observed wavelength of \NIII\ (denoted by dotted lines) 
between the data and the pipeline models (that do not account for \NIII ). 
The SDSS automatic pipeline provides a poor fit for \obj\ largely due to the significant 
\NIII\ emission, which is not being accounted for in the template (because this feature is much weaker for normal quasars). 
All spectra (not model) have been smoothed by a 4-pixel boxcar for better clarity.
Spectra for the comparison fibers are offset vertically for display purposes.
\label{fig:test_plate}
}
\end{figure*}

Finally, we have examined the co-added SDSS spectra taken on the same plate at the same time but with other fibers to rule out artifacts such as those due to contamination from residual sky lines, the high pressure sodium (HPS) lamps \citep{Law2016}, and/or SDSS pipeline reduction issues. In no case do we detect a residual emission as significant as in \obj\ at the position of N III] after subtracting the continuum.  Figure \ref{fig:test_plate} shows six such examples (for fibers 320, 325, 329, 330, 335, and 357, which were randomly chosen after removing the ones that happen to have other emission lines at the position of \NIII , such as fibers 326, 331, and 332).  The \NIII\ is close in wavelengths to the HPS lamps, which caused increased noise levels as seen in the larger error bars in Figure \ref{fig:spec_wfit}. But the fact that \NIII\ is blueshifted with respect to the HPS feature argues against it as the culprit.


\end{appendices}

\bibliography{/Users/zeus/Documents/References/binaryrefs}

\end{document}